\documentclass[12pt]{article}

\renewcommand{\arraystretch}{1.2}

\usepackage{caption}
\usepackage[utf8]{inputenc}
\usepackage{multirow}
\usepackage[pdftex]{graphicx}
\usepackage{amssymb}
\usepackage{amsmath}
\usepackage{cite}
\usepackage{braket}
\usepackage{longtable}
\usepackage{booktabs}
\usepackage{array}
\usepackage[normalem]{ulem}
\usepackage[dvipsnames]{xcolor}
\usepackage{xspace}
\usepackage[]{hyperref}

\newcommand{\nn}{\nonumber}
\newcommand{\schii}[2]{\chi^{(#1)}_{#2}}
\newcommand{\schi}[3]{\chi^{(#1)}_{#2}\big|_{#3}}

\newcommand{\A}{{\cal A}}

\newcommand{\F}{{\cal F}}

\newcommand{\N}{{\cal N}} 
\renewcommand{\P}{{\cal P}} 
\renewcommand{\S}{{\cal S}} 
\newcommand{\T}[1]{\mathcal{T}\left\lbrace #1 \right\rbrace}

\renewcommand{\Im}{\text{Im}}
\newcommand{\eps}{\varepsilon}

\newcommand{\lamkin}{\lambda_{\text{kin}}}
\newcommand{\FM}[2][B\to M]{\ensuremath{\F^{#1}_{#2}}}
\newcommand{\hFM}[2][B\to M]{\ensuremath{\hat{\F}^{#1}_{#2}}}

\newcommand{\Eqs}[2]{Eqs.~\ref{#1}-\ref{#2}}

\newcommand{\Reff}[1]{Ref.~\cite{#1}}

\newcommand{\ie}{i.e.\xspace}

\numberwithin{equation}{section}
\makeatletter
\let\oldtheequation\theequation
\renewcommand\tagform@[1]{\maketag@@@{\ignorespaces#1\unskip\@@italiccorr}}
\renewcommand\theequation{(\oldtheequation)}
\makeatother

\newcommand{\EOS}{\texttt{EOS}\xspace}

\newcounter{TODO}

\allowdisplaybreaks

\begin{document}

\begin{flushright}
EOS-2023-02\\
IPPP/23/22\\
P3H-23-026\\
SI-HEP-2023-09\\
\end{flushright}

$\ $
\vspace{-2mm}
\begin{center}
\fontsize{15}{20}\selectfont
\bf
\boldmath
Dispersive Analysis of $B\to K^{(*)}$ and $B_s\to \phi$ Form Factors
\end{center}

\vspace{2mm}

\begin{center}
{Nico Gubernari$^{\,a}$, M\'eril Reboud$^{\,b}$, Danny van Dyk$^{\,b}$, Javier Virto$^{\,c,d}$}\\[5mm]
{\it\small
{$^{\, a}$} 
Universit\"at Siegen, Naturwissenschaftliche Fakult\"at,\\
Walter-Flex-Stra\ss{}e 3, 57068 Siegen, Germany
\\[2mm]
{$^{\, b}$} 
Institute for Particle Physics Phenomenology and Department of Physics,\\
Durham University, Durham DH1 3LE, UK
\\[2mm]
{$^{\, c}$}
Departament de Física Quàntica i Astrofísica,
Universitat de Barcelona,\\
Martí i Franqués 1, 08028 Barcelona, Catalunya
\\[2mm]
{$^{\, d}$}
Institut de Ciències del Cosmos (ICCUB), Universitat de Barcelona,\\
Martí i Franqués 1, 08028 Barcelona, Catalunya
\\[2mm]
}
\end{center}

\vspace{1mm}
\begin{abstract}\noindent
\vspace{-5mm}

\noindent
We perform the first simultaneous dispersive analysis of the $B\to K$, $B\to K^*$, and $B_s\to \phi$ form factors.
By means of an improved parametrization, we take into account the form factors' below-threshold branch cuts arising from on-shell $\bar{B}_s \pi^0$ and $\bar{B}_s \pi^0 \pi^0$  states,
which so far have been ignored in the literature.
In this way, we eliminate a source of hard-to-quantify systematic uncertainties.
We provide machine-readable files to obtain the full set of the $\bar{B}\to \bar{K}^{(*)}$ and $\bar{B}_s\to \phi$ form factors
in and beyond the entire semileptonic phase space. 
\end{abstract}

\newpage

\tableofcontents

\section{Introduction}

Rare $B$ decays are decays of $B$ mesons with small branching ratios of the order of $10^{-6}$ or less. In the Standard Model (SM) these tiny rates are explained by the fact that they are loop-mediated processes with CKM or GIM suppression.
The fact that experimental measurements of these rates are of the same order magnitude of the SM predictions already poses strong indirect constraints on Beyond-the-SM (BSM) physics, most prominently in models where these processes arise at tree level. 
This ability to pose strong constraints on BSM physics is the same as discovery power~\cite{Isidori:2010kg}.

A class of rare $B$ decays that has been discussed extensively in over ten years of LHC operations are the Flavour-Changing Neutral Current (FCNC) exclusive processes of the type $b\to s \ell^+\ell^-$, such as $\bar{B}\to \bar{K}\ell^+\ell^-$, $\bar{B} \to \bar{K}^*\ell^+\ell^-$, and $\bar{B}_s \to \phi\ell^+\ell^-$. 
While the importance of these measurements was clear and had been discussed before~\cite{Bobeth:2007dw,Altmannshofer:2008dz,Bobeth:2010wg,Matias:2012xw,Beaujean:2012uj,Descotes-Genon:2012isb}, a generalised interest was driven by the ``anomalies'' observed starting from 2013 \cite{LHCb:2013ghj,Descotes-Genon:2013wba,Altmannshofer:2013foa,Beaujean:2013soa,Horgan:2013pva,Hurth:2013ssa,Altmannshofer:2014rta,LHCb:2014vgu,Descotes-Genon:2015uva,Ciuchini:2015qxb,LHCb:2015svh,LHCb:2017avl}. 
While the deviations in the observables measuring lepton-flavour non-universality seem to be now more in agreement with the SM, the deviations in the muonic observables are still puzzling. Independently of this, the original motivations to study these rare decays have only been strengthened by the sheer amount and the precision of the experimental measurements. These measurements will clearly be one of the legacies of the LHC era.
\\

On the theoretical side, efforts are focused on the control of hadronic uncertainties that are crucial to the unambiguous interpretation of the data. All such exclusive $b\to s \ell^+\ell^-$ observables can be computed from a handful of decay amplitudes given by (see, e.g.,~\cite{Gubernari:2022hxn})
\begin{equation}
\A_\lambda^{L,R} = \N \bigg[
(C_9 \mp C_{10}) {\F}_\lambda(q^2) + \frac{2m_b M_B}{q^2} \Big\{  C_7 {\F}_{T,\lambda}(q^2) - 16\pi^2 \frac{M_B}{m_b} {\cal H}_\lambda(q^2)  \Big\}
\bigg] + {\cal O}(\alpha_\text{em}).
\end{equation}
Here, $q$ is the four-momentum of the lepton pair,
$L/R$ indicates the lepton chirality, $\lambda$ is the dilepton polarisation, $C_i$ are Wilson coefficients of the Effective Field Theory (EFT) below the weak scale, and $\F_{(T),\lambda}(q^2)$, ${\cal H}_\lambda(q^2)$ are respectively \emph{local} and \emph{non-local} hadronic form factors (FFs). These hadronic FFs depend on the initial and final states and on QCD infrared scales, and cannot be calculated in perturbation theory.
Both types of FFs have been the subject of intensive research \cite{Ball:1998kk,Charles:1998dr,Beneke:2000wa,Bauer:2000yr,Beneke:2003pa,Ball:2004ye,Khodjamirian:2006st,Duplancic:2008ix,Jager:2012uw,Horgan:2013pva,Bharucha:2015bzk,Gubernari:2018wyi,Descotes-Genon:2019bud,Parrott:2022rgu,Descotes-Genon:2023ukb,Aoki:2021kgd,Beneke:2001at,Grinstein:2004vb,Beylich:2011aq,Khodjamirian:2010vf,Khodjamirian:2012rm,Bobeth:2017vxj,Arbey:2018ics,Asatrian:2019kbk,Gubernari:2020eft}. Although the non-local type is considerably more complicated, they can be written in terms of local FFs at leading power in an operator product expansion (OPE)~\cite{Beneke:2001at,Grinstein:2004vb,Khodjamirian:2010vf,Asatrian:2019kbk}.
Hence, understanding the local FFs precisely seems a more urgent matter.

The FFs $\F_{(T),\lambda}$ are functions of the momentum transfer $q^2$ and different determinations of the FFs apply at different values of $q^2$. For instance, determinations based on Light-Cone Sum Rules (LCSRs) apply at low values of $q^2$, while those based on Lattice QCD (LQCD) calculations have (until recently) only applied to the high end of the $q^2$ spectrum. Combining all the known information on the FFs in order to be able to predict observables in \emph{any} given region of $q^2$ thus requires to understand the $q^2$ dependence. Fortunately, this $q^2$ dependence is simpler than the normalisation itself, and rigorous parametrizations of the $q^2$ dependence of the FFs can be derived from the fundamental principles of analyticity and unitarity~\cite{Okubo:1971jf}.

These parametrizations are based on Taylor expansions of analytic functions, and while rigorous, in real applications they must be truncated in order to feature a finite number of parameters to be fitted. This truncation introduces a systematic uncertainty that is difficult to assess.
Fortunately, one can derive absolute bounds on some combinations of FFs integrated over a kinematic region, by relating the exclusive to the inclusive rates and making use of a dispersion relation. These bounds are called ``dispersive'' or ``unitarity'' bounds, and effectively constrain the truncation error in specific parametrizations.
The dispersive bounds have been used within the so-called BGL and BCL parametrizations~\cite{Boyd:1994tt,Boyd:1997kz,Bourrely:2008za}, and have been proven extremely useful in $\bar{B}\to D^{(*)}$~\cite{Bigi:2017jbd} and $\Lambda_b\to \Lambda^{(*)}$ semileptonic transitions~\cite{Blake:2022vfl, Amhis:2022vcd}.

The purpose of this paper is to revisit the parametrization for $\bar{B}\to \bar{P}$ and $\bar{B}\to \bar{V}$ FFs. 
We shall demonstrate that the parametrization of~\Reff{Gubernari:2020eft} is advantageous when below-threshold branch cuts appear.
We will also apply these improvement to carry out the first simultaneous dispersive analysis of all $\bar{B}\to K^{(*)}$ and $\bar{B}_s \to \phi$ FFs. 

We begin in \autoref{sec:th-fr} with a detailed discussion of the theoretical framework, first defining the FFs and the parametrizations, and then discussing the dispersive bounds.  
In \autoref{sec:analysis} we present a complete numerical analysis where we fit the to the parameters of our FF parametrization including the improved dispersive bounds. Our conclusions are presented in \autoref{sec:conclusions}.
Supplementary material and a discussion of the polynomials used in the parametrization are provided in \autoref{app:supplementary_material}
and \autoref{app:orthonormal_polynomials}.

\section{Theoretical framework}
\label{sec:th-fr}

In this section we provide the theoretical framework that we use to perform our analysis of \autoref{sec:analysis}.
In \autoref{sec:ff-def}, we define the $\bar{B} \to \bar{M}$ form factors (FFs) starting from the corresponding matrix elements.
Then, we briefly review the FF parametrizations written in terms of the conformal variable usually denoted by $z$, which have been commonly used in the literature.
In \autoref{sec:th:bounds}, we introduce the dispersive bounds, which constrain the parameters of certain $z$ parametrizations.
In \autoref{sec:th:param}, we give an accurate description of the FF parametrization that we adopt, which is based on the parametrization of Ref.~\cite{Gubernari:2020eft} and allows for a straightforward implementation of the dispersive bounds.

\subsection[Form factor definitions and z parametrizations]{
    Form factor definitions and $\boldsymbol{z}$ parametrizations
}
\label{sec:ff-def}

Hadronic FFs are scalar functions of the momentum transfer squared $q^2=(p-k)^2$ that arise in the Lorentz decomposition of exclusive hadronic matrix elements.
We define the $\bar{B} \to \bar{P}$(seudoscalar) or $\bar{B} \to \bar{V}$(ector) meson FFs following the same notation as used in Ref.~\cite{Gubernari:2018wyi}, which
relies on a commonly-used Lorentz decomposition.
In this work we focus on the $\bar{B} \to \bar{K}$, $\bar{B} \to \bar{K}^*$, and $\bar{B}_s \to \phi$ transitions --- which we collectively denote as $\bar{B} \to \bar{M}$ --- although both the definitions given below and our results of \autoref{sec:th:bounds} and \autoref{sec:th:param} can be used for any $\bar{B} \to \bar{P},\bar{V}$ transition.
Note that we do not make any attempt here to account for the instability of the vector meson~\cite{Descotes-Genon:2019bud}.

The three local $\bar{B}\to \bar{P}$ FFs are defined by 
\begin{align}
    \langle\bar  P(k) | J_V^\mu | \bar{B}(p) \rangle &=
        \left[ (p + k)^\mu - \frac{M_B^2 - M_P^2}{q^2} q^\mu \right] f_+^{B\to P}
        + \frac{M_B^2 - M_P^2}{q^2} q^\mu f_0^{B\to P}, \\
    \langle\bar  P(k) | J_T^\mu | \bar{B}(p) \rangle &=
        \frac{i f_T^{B\to P}}{M_B + M_P} \left[ q^2 (p + k)^\mu - (M_B^2 - M_P^2) q^\mu \right].
\end{align}
The seven local $\bar{B}\to \bar{V}$ FFs are defined by
\begin{align}
    \langle\bar  V(k, \eta) | J_V^\mu |\bar{B}(p) \rangle &=
        \epsilon^{\mu\nu\rho\sigma} \eta_\nu^* p_\rho k_\sigma \frac{2 V^{B\to V}}{M_B + M_V}, \\
    \label{eq:th:param:A}
    \langle\bar  V(k, \eta) | J_A^\mu |\bar{B}(p) \rangle &=
        i \eta_\nu^* \bigg[ g^{\mu\nu} (M_B + M_V) A_1^{B\to V}
        - (p + k)^\mu q^\nu  \frac{A_2^{B\to V}}{M_B + M_V} \nn \\
        & \hspace{15mm} - 2 M_V \frac{q^\mu q^\nu}{q^2} (A_3^{B\to V} - A_0^{B\to V}) \bigg],\\
    \langle\bar  V(k, \eta) | J_T^\mu | \bar{B}(p) \rangle &=
        \epsilon^{\mu\nu\rho\sigma} \eta_\nu^* p_\rho k_\sigma \, 2 T_1^{B\to V}, \\
    \langle\bar  V(k, \eta) | J_{AT}^\mu |\bar{B}(p) \rangle &=
        i \eta_\nu^* \bigg[ \Big( g^{\mu\nu} (M_B^2 - M_V^2) - (p + k)^\mu q^\nu \Big) T_2^{B\to V} \nn \\
        & \hspace{15mm} + q^\nu \left( q^\mu - \frac{q^2}{M_B^2 - M_V^2} (p + k)^\mu \right) T_3^{B\to V} \bigg],
\end{align}
where $\eta$ is the polarisation four-vector of the vector meson, and we abbreviate
\begin{equation}
    A_3^{B\to V} \equiv \frac{M_B + M_V}{2 \, M_V} \, A_1^{B\to V} - \frac{M_B - M_V}{2 \, M_V} \, A_2^{B\to V}.
\end{equation}
The local currents $J_\Gamma^\mu$ used above read
\begin{equation}
\begin{aligned}
    \label{eq:th:ff-def:JGamma}
    J_V^\mu & = \bar{s}\, \gamma^\mu b    \,,&\qquad
    J_A^\mu & = \bar{s}\, \gamma^\mu \gamma_5 b      \,,\\
    J_T^\mu & = \bar{s}\, \sigma^{\mu \alpha}q_\alpha  b   \,,&
    J_{AT}^\mu & = \bar{s}\, \sigma^{\mu \alpha}q_\alpha \gamma_5  b
    \,.
\end{aligned}
\end{equation}
Throughout this work, we keep the $q^2$ dependence of the FFs implicit unless necessary.
In order to diagonalize the dispersive bounds (see \autoref{sec:th:bounds}), we introduce the helicity FFs
\begin{align}
    A_{12}^{B\to V} &= \frac{(M_B + M_V)^2 (M_B^2 - M_V^2 - q^2) \, A_1^{B\to V} - \lamkin \, A_2^{B\to V}}{16 \, M_B M_V^2 (M_B + M_V)}\,, \\
    \label{eq:th:T23}
    T_{23}^{B\to V} &= \frac{(M_B^2 - M_V^2) (M_B^2 + 3 M_V^2 - q^2) \, T_2^{B\to V} -  \lamkin \, T_3^{B\to V}}{8 \, M_B M_V^2 (M_B - M_V)}\,,
\end{align}
where $ \lamkin \equiv \lamkin(q^2) \equiv \lambda(M_B^2,M_M^2,q^2)=(M_B^2 - M_M^2 -q^2)^2-4M_M^2q^2$ is the Källén function.
The absence of dynamic singularities at $q^2 = 0$ and the antisymmetry of $\sigma^{\mu\nu}$ under the exchange $\mu \leftrightarrow \nu$ imply the following relations
\begin{equation}
    f_+^{B\to P}(0) = f_0^{B\to P}(0), \quad
    A_0^{B\to V}(0) = A_3^{B\to V}(0), \quad
    T_1^{B\to V}(0) = T_2^{B\to V}(0)\,.
\end{equation}
Due to the symmetries of the helicity amplitudes at the kinematical endpoint $q^2 = s_- \equiv (M_B - M_V)^2$, there are two additional constraints on the FFs~\cite{Hiller:2013cza}: 
\begin{align}
    A_{12}^{B\to V}(s_-) &= \frac{(M_B + M_V)(M_B^2 - M_V^2 - s_-)}{16 \, M_B M_V^2} \, A_1^{B\to V}(s_-)\,, \\
    T_{23}^{B\to V}(s_-) &= \frac{(M_B + M_V)(M_B^2 + 3 M_V^2 - s_-)}{8 \, M_B M_V^2} \, T_2^{B\to V}(s_-)\,.
\end{align}

\bigskip

The FFs are sensitive to infrared QCD scales and thus cannot be directly calculated in perturbation theory.
Lattice QCD (LQCD) provides a means to perform such non-perturbative calculations.
However, most of the presently available LQCD calculations for the FFs provide results only in the high-$q^2$ region of the semileptonic phase-space.
Therefore, one needs to either extrapolate the LQCD results to the low-$q^2$ region or to combine them with results obtained at low-$q^2$ using other methods.
In both cases, a suitable parametrization of the FFs is needed.
In the absence of LQCD data at all or in parts of the phase space, an alternative QCD-based method is the framework of light-cone sum rules (LCSRs).
The LCSRs provide estimates of the FFs at low $q^2$, which can be used to anchor their parametrizations.
Their use incurs an irreducible systematic uncertainty due to the modelling of the corresponding distribution amplitudes and the use of semi-global quark-hadron duality~\cite{Shifman:2000jv}.

The most frequently-used parametrizations in the literature describing the $\bar{B} \to \bar{M}$ FFs are the Boyd-Grinstein-Lebed (BGL)~\cite{Boyd:1994tt,Boyd:1997kz}, the Bourrely-Caprini-Lellouch (BCL)~\cite{Bourrely:2008za}, and the Bharucha-Straub-Zwicky (BSZ)~\cite{Bharucha:2015bzk}  parametrizations.
Even though the BSZ parametrization was proposed more than thirty years later than the BGL one and is inspired by it, the former has a much simpler form than the latter since it is not devised to implement the dispersive bounds.
For this reason, we first review the BSZ parametrization in this section in order to introduce the relevant notation, while we discuss the BGL parametrization in \autoref{sec:th:bounds} and \autoref{sec:th:param}. 

To understand how these parametrizations have been introduced, one first needs to understand the analytical structure of the $\bar{B} \to \bar{M}$ FFs. 
Due to on-shell particle production, a FF arising from a specific current $J_\Gamma^\mu$ contains a number of simple poles from QCD-stable one-particle states, and develops a branch cut on the positive real axis for $q^2 \geq s_\Gamma$, where $s_\Gamma$ is the first multi-particle threshold.
The dispersively bounded parametrizations BGL and BCL are only valid in the particular case where $s_\Gamma$ coincides with $s_+ \equiv (M_B + M_M)^2$.
We note that $s_\Gamma = s_+$ holds for processes like $\bar{B}\to\pi$ or $D\to \pi$, but it does \emph{not} hold for the processes considered in this work: $\bar{B} \to \bar{K}$, $\bar{B} \to \bar{K}^*$, and $\bar{B}_s \to \phi$.
In fact, the vector and tensor $\bar{B} \to \bar{M}$ FFs have $s_V = s_T = (M_{B_s} + M_{\pi^0})^2$ and the axial and axial-tensor  $\bar{B} \to \bar{M}$ FFs have
$s_A = s_{AT} = (M_{B_s} + 2 M_{\pi^0})^2$.
In order to deal with the case $s_\Gamma < s_+$, we use the parametrization first proposed in Ref.~\cite{Gubernari:2020eft}, which we review in \autoref{sec:th:param}.
A recent alternative way to achieve the application of the dispersive bounds when $s_\Gamma < s_+$ is discussed in Ref.~[54] for $B_s \to K$ transitions.
The improvements to the dispersive bounds presented in this work can be straightforwardly applied to the approach of Ref.~[54].

\begin{figure}[t!]
    \centering
    \includegraphics[width=.95\textwidth]{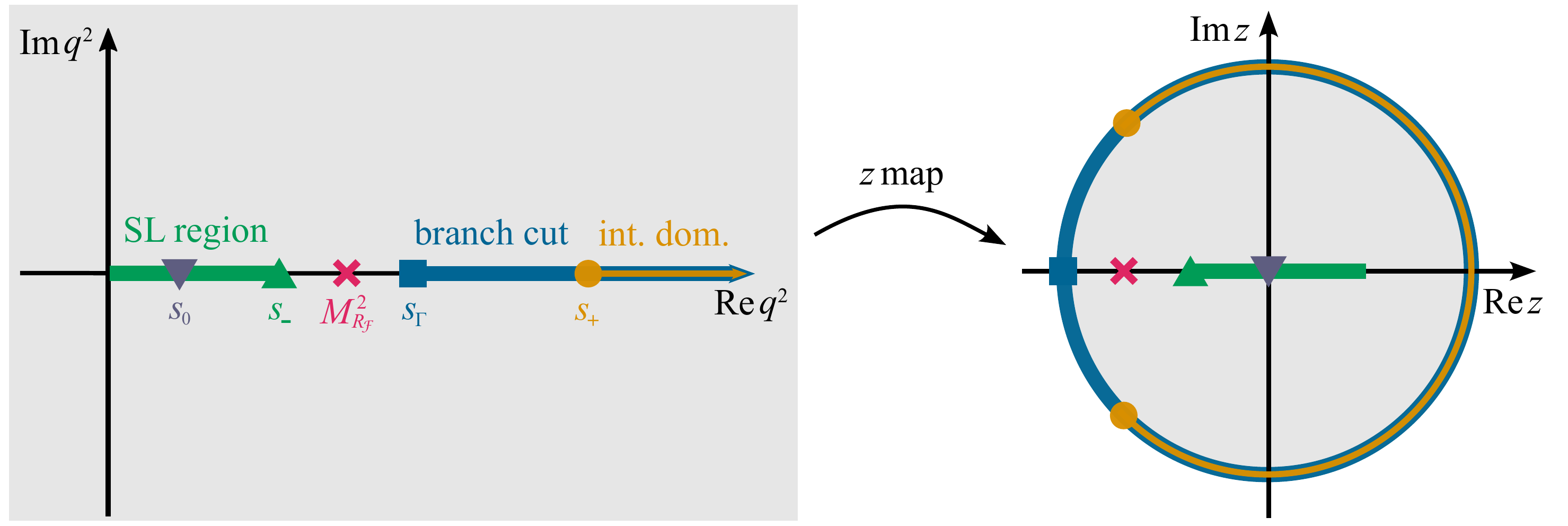}
    \caption{%
        Illustration of the relevant regions in momentum space of the FFs both in the $q^2$ and the $z$ plane.
        The semileptonic (SL) region extends from $q^2=0$ to $q^2=s_-\equiv(M_B - M_M)^2$, the branch cut starts at $q^2=s_\Gamma$, while the region for $q^2 \geq s_+$ is the one relevant for the dispersive bounds (see \autoref{sec:th:bounds}).
    }
    \label{fig:z_map}
\end{figure}

Following these considerations, we use the conformal mapping
\begin{equation}
    \label{eq:zmap}
    q^2 \mapsto z(q^2) \equiv z(q^2; s_\Gamma, s_0) = 
    \frac{
        \sqrt{s_\Gamma-q^2}-\sqrt{s_\Gamma-s_0^{\phantom{2}}}
    }{
        \sqrt{s_\Gamma-q^2}+\sqrt{s_\Gamma-s_0^{\phantom{2}}}
    }\,,
\end{equation}
where $s_0$ is a free parameter that can be chosen in the interval $(-\infty,s_\Gamma)$.
It is convenient to set $s_0 = (M_B + M_M)\left(\sqrt{M_B} - \sqrt{M_M}\right)^2$, which minimises $|z|$ in the semileptonic phase space. 
The transformation \autoref{eq:zmap} maps the complex $q^2$ plane onto the complex $z$ plane. 
In particular, the first Riemann sheet of the complex $q^2$ plane is mapped
onto the open unit disk centred around the origin $z=0$,
and the second Riemann sheet is mapped onto the complement of the closed unit disk.
The branch cut connecting both sheets is mapped onto the unit circle.
We provide an illustration in \autoref{fig:z_map}.

The FFs are analytic functions of $z$ on the open unit disk, except for a finite number of simple poles due to the aforementioned on-shell one-particle states, and which can be easily accounted for (see \autoref{sec:th:param}).
Once the poles are removed, the resulting FFs can be expanded in a Taylor series that converges for $|z|<1$. 
This is the essence of all $z$ parametrizations, including BCL, BGL and BSZ. The latter parametrization reads
\begin{align}
    \label{eq:BSZparam}
    \FM{\Gamma,\lambda} (q^2)
    =
    \frac{1}{1 - \frac{q^2}{M_\F^2}}
    \sum_{n=0}^\infty
    c_n^\F
    \left[
        z(q^2)
        -
        z(0)
    \right]^n
    \,,
\end{align}
where $\FM{\Gamma,\lambda}$ is a $\bar{B} \to \bar{M}$ FF and $M_\F$ is the mass of the lowest-lying one-particle state with same spin, parity, and flavour quantum numbers as $\FM{\Gamma,\lambda}$.
These masses can be found in Tab.~3 of Ref.~\cite{Bharucha:2015bzk}.

The simple form of \autoref{eq:BSZparam} makes this parametrization convenient and its convergence in the semileptonic region is ensured by the analytic properties of the FFs.
In any phenomenological application, the parametrization \eqref{eq:BSZparam} would
be truncated at some order $n=N$.
However, it is not readily possible to estimate the truncation error,
that is to estimate the uncertainty due to the truncated terms with
coefficients $c_n$ for $n>N$.
To estimate the truncation error, one needs to use additional constraints. 
These constraints can be obtained using unitarity and analyticity, and are usually called \emph{dispersive (or unitarity) bounds}.
In \autoref{sec:th:bounds} and \autoref{sec:th:param} we review the derivation of these constraints.

\subsection{Dispersive bounds}
\label{sec:th:bounds}

The application of dispersive bounds is commonplace for charged-current heavy-to-heavy transitions, such as
$\bar{B}\to D^{(*)}$~\cite{Boyd:1997kz,Caprini:1997mu,Bordone:2019vic,Bordone:2019guc}.
Their application to the local FFs in $\bar{B}\to \bar{K}^{(*)}$ transitions
has been discussed for the first time in Ref.~\cite{Bharucha:2010im}. 
In this work, we improve upon this analysis in two ways.
First, by analysing simultaneously the $\bar{B} \to \bar{K}$, $\bar{B} \to \bar{K}^*$, and $\bar{B}_s \to \phi$ transitions
(as discussed in this section),
and second, by accounting for subthreshold branch cuts that are not included in Ref.~\cite{Bharucha:2010im} (as discussed in
\autoref{sec:th:param}).
To derive dispersive bounds for any of the $\bar{B} \to \bar{M}$ FFs, one commonly uses the two-point correlation functions 
\begin{align}
    \label{eq:th:bounds:correlator}
    \Pi^{\mu\nu}_{\Gamma}(q)
        & \equiv i\! \int\! d^4x\, e^{iq\cdot x} \braket{0 | \T {J_\Gamma^\mu(x) J_\Gamma^{\dagger,\nu}(0)} | 0}\,,
\end{align}
where the currents $J_\Gamma^\mu$ have been defined in \autoref{eq:th:ff-def:JGamma}.
It is convenient to decompose the tensor-valued correlator $\Pi^{\mu\nu}_{\Gamma}(q)$ in terms
of scalar-valued functions $\Pi_\Gamma^{(J)}(q^2)$,
\begin{align}
    \label{eq:th:bounds:generic}
    \Pi^{\mu\nu}_{\Gamma}(q)
        \equiv \sum_{J} \S^{\mu\nu}_{(J)} (q) \, \Pi_\Gamma^{(J)}(q^2)\ .
\end{align}
Our choice of decomposition involves structures with definite total angular momentum $J$:
\begin{align}
    \label{eq:th:bounds:common}
    \S^{\mu\nu}_{(J=1)}(q) & = \left(\frac{q^\mu q^\nu}{q^2} - g^{\mu\nu}\right)\,, &
    \S^{\mu\nu}_{(J=0)}(q) & = \frac{q^\mu q^\nu}{q^2}\,.
\end{align}
The corresponding projectors read $\mathcal{P}^{\mu\nu}_{(J=1)} = \S^{\mu\nu}_{(J=1)} / (d - 1)$
and $\mathcal{P}^{\mu\nu}_{(J=0)} = \S^{\mu\nu}_{(J=0)}$, where $d$ is the number of spacetime dimensions.
This decomposition has been frequently used in the literature~\cite{Boyd:1997kz,Caprini:1997mu,Bharucha:2010im}.
Its connection to the total angular momentum $J$ provides insights into the FFs'
properties.

The discussion so far is only helpful if a connection can be made between the scalar functions $\Pi_\Gamma^{(J)}$ as defined in \autoref{eq:th:bounds:generic} and the hadronic FFs.
Key to this connection is that the former fulfil a subtracted dispersion relation:
\begin{align}
    \label{eq:th:bounds:chisub}
    \schii{J}{\Gamma}(Q^2)
        & = \frac{1}{n!} \left[\frac{\partial}{\partial q^2}\right]^n \Pi_\Gamma^{(J)}(q^2)
        \bigg|_{q^2=Q^2}
          = \frac{1}{\pi} \int\limits_0^\infty ds\, \frac{\Im \Pi_\Gamma^{(J)}(s)}{(s - Q^2)^{n+1}}\,.
\end{align}
Here $Q^2$ is the subtraction point and $n$ is the number of subtractions, which is chosen \emph{a posteriori} such that the integral on the r.h.s.~of \autoref{eq:th:bounds:chisub} converges. 
On the one hand, the discontinuities of $\Pi_\Gamma^{(J)}$ are known from a computation within a local OPE,
and the results are valid for $Q^2 \ll (m_b + m_s)^2$. 
We use the results provided in Ref.~\cite{Bharucha:2010im}, which include
perturbative corrections up to next-to-leading order in $\alpha_s$ for the dimension-$0$ term
and power corrections due to operators up to dimension $5$.
For convenience, we quote the numerical results of Ref.~\cite{Bharucha:2010im} in \autoref{tab:resonances},
alongside further information.

\begin{table}
    \centering
    \renewcommand{\arraystretch}{1.4}
    \setlength{\tabcolsep}{5pt}
    \begin{tabular}{@{}l c c c@{}}
        \toprule
        $\schii{J}{\Gamma}$
                                   & $\schi{J}{\Gamma}{\mathrm{OPE}}\times 10^2$
                                                      & Form factor                                   & Pole $R_\F$ (mass, decay constant) $[\mathrm{GeV}]$ \\
        \midrule
        $\schii{J=0}{V}$             & $1.42$         & $f_0^{B\to K}$                                & --- \\
        $\schii{J=0}{A}$             & $1.57$         & $A_0^{B\to K^*}, A_0^{B_s\to\phi}$            & $\bar{B}_s  \, (5.367, 0.2307(13))$ \\
        $\schii{J=1}{V}$             & $1.20/m_b^2$   & $f_+^{B\to K}, V^{B\to K^*}, V^{B_s\to\phi}$  & $\bar{B}_s ^* \, (5.416, 0.2231(56))$ \\
        $\schii{J=1}{A}$             & $1.13/m_b^2$   & $A_{12}^{B\to K^*}, A_{12}^{B_s\to\phi}, A_1^{B\to K^*}, A_1^{B_s\to\phi}$
                                                                                                      & --- \\
        $\schii{J=1}{T}$             & $0.803/m_b^2$  & $f_T^{B\to K}, T_1^{B\to K^*}, T_1^{B_s\to\phi}$
                                                                                                      & $\bar{B}_s ^* \, (5.416, 0.236(22))$ \\
        $\schii{J=1}{AT}$            & $0.748/m_b^2$  & $T_{23}^{B\to K^*}, T_{23}^{B_s\to\phi}, T_2^{B\to K^*}, T_2^{B_s\to\phi}$
                                                                                                      & --- \\
        \bottomrule
    \end{tabular}
    \caption{
        \label{tab:resonances}
        List of the functions $\schii{J}{\Gamma}$ considered in this work with the corresponding OPE results (taken from Ref.~\cite{Bharucha:2010im}), affected FFs, and sub-threshold poles. 
        The reference value for the $b$-quark mass is $m_b=4.2\,\text{GeV}$.
        The parameters of the poles are taken from Refs.~\cite{ParticleDataGroup:2020ssz,Lang:2015hza,Lubicz:2017asp,Pullin:2021ebn,Bazavov:2017lyh}.
    }
\end{table}

On the other hand, unitarity of the $S$-matrix implies that the imaginary part of the functions $\Pi_\Gamma^{(J)}$ can be expressed  as an infinite sum of
exclusive hadronic matrix elements of the currents $J_\Gamma$ (see, e.g., Refs.~\cite{Boyd:1997kz,Caprini:1997mu,Bharucha:2010im}):
\begin{multline}
    \label{eq:ImPi}
    \Im\,\Pi_\Gamma^{(J)}(s + i \eps) 
        = \frac{1}{2} \sum \!\!\!\!\!\!\!\! \int\limits_H d\rho_H (2\pi)^4 \delta^{(4)}(p_H - q)
        \mathcal{P}_{\mu\nu}^{(J)}
            \braket{0 | J_\Gamma^\mu | H(q)}\braket{\bar H(q) | J_\Gamma^{\dagger,\nu} \!| 0}\Big|_{q^2=s} 
    .
\end{multline}
Here we sum over all hadronic states $H$ with flavour quantum numbers $S = -B = 1$ and we integrate over the corresponding phase space.
For every state $H$, we can isolate a positive semidefinite exclusive contribution to $\schii{J}{\Gamma}$, which we denote as $\schi{J}{\Gamma}{H}$.
For one-particle bound states $H = \bar{B}_s^{(*)}$, the contributions read
\begin{equation}
\begin{aligned} 
    \label{eq:1ptcontr}
    &
    \schi{J=1}{V}{\mathrm{1pt}} = \frac{M_{B_s^*}^2 f_{B_{s}^*}^2}{(M_{B_s^*}^2-Q^2)^3}\,,  
    &&\quad
    \schi{J=0}{A}{\mathrm{1pt}} = \frac{M_{B_s}^2 f_{B_s}^2}{(M_{B_s}^2-Q^2)^2}\,, &
    &&\quad
    \schi{J=1}{T}{\mathrm{1pt}} = \frac{M_{B_s^*}^4 (f_{B_s^*}^T)^2}{(M_{B_s^*}^2-Q^2)^4}\,. & 
\end{aligned}
\end{equation}
The decay constants are defined as
\begin{equation*}
\begin{aligned}
    \braket{0 | J_V^\mu | \bar{B}_s^*(q,\eta)}
        & = i   M_{B_s^*} f_{B_s^*} \eta^\mu,
    &
    \braket{0 | J_A^\mu | \bar{B}_s(q)}
        & = i q^\mu f_{B_s},
    &
    \braket{0 | J_T^\mu | \bar{B}_s^*(q,\eta)}
        & = i M_{B_s^*}^2 f_{B_s^*}^T \eta^\mu,
\end{aligned}
\end{equation*}
where $\eta$ is the polarisation four-vector of the $\bar{B}_s^*$ meson.
Some comments are in order: we do not include all known one-particle $b\bar{s}$ states.
Rather, we consider only bound states, i.e., only states with masses below the current-specific
threshold $s_\Gamma$.
This is necessary, since $b\bar{s}$ states above their respective threshold(s) are \emph{resonances}
within the two-particle spectrum. As such, they emerge on the FFs' second (or higher) Riemann sheets
rather than the first sheets.
As a consequence, the resonances' contributions are accounted for by contributions of the various
FFs, both two-particle FFs and those for higher multiplicities.
In this way, we avoid double counting of the resonances' contributions.

If $H$ is a two-particle state, the $H$-to-vacuum matrix elements can be related through crossing symmetry to hadronic FFs.
For this work, we restrict our analysis to the states $H=\bar{B}M \equiv\bar  BK,\bar{B}K^*,\bar{B}_s \phi$.
We obtain the exclusive contributions $\schi{J}{\Gamma}{H}$ following the derivation of Ref.~\cite{Boyd:1997kz}.
For $H=\bar{B}K$ and in the common choice of form factor basis, these results read
\begin{align}
\label{eq:chi_time_V}
    \schi{J=0}{V}{\bar{B}K}  =&\, \frac{\eta^{B\to K}}{16 \pi^2} \int\limits_{(M_B + M_K)^2}^\infty ds \frac{ \lamkin^{1/2}(s)}{s^2 (s - Q^2)^2}
        \left(M_B^2 - M_K^2\right)^2 \, |f_0^{B\to K}(s)|^2 \,, \\
    \schi{J=1}{V}{\bar{B}K}  =&\, \frac{\eta^{B\to K}}{48 \pi^2} \int\limits_{(M_B + M_K)^2}^\infty ds \frac{ \lamkin^{3/2}(s)}{s^2 (s - Q^2)^3}
        \, |f_+^{B\to K}(s)|^2 \,, \\
    \schi{J=1}{T}{\bar{B}K}  =&\, \frac{\eta^{B\to K}}{48 \pi^2} \int\limits_{(M_B + M_K)^2}^\infty ds \frac{ \lamkin^{3/2}(s)}{(s - Q^2)^4}
        \frac{1}{(M_B + M_K)^2} \, |f_T^{B\to K}(s)|^2 \,.
\end{align}
For $H=\bar{B}K^*$, they read
\begin{align}
    \schi{J=1}{V}{\bar{B}K^*}  =&\, \frac{\eta^{B\to K^*}}{24 \pi^2} \int\limits_{(M_B + M_{K^*})^2}^\infty ds \frac{ \lamkin^{3/2}(s)}{s (s - Q^2)^3}
        \frac{1}{(M_B + M_{K^*})^2} \, |V^{B\to K^*}(s)|^2 , \\
    \schi{J=0}{A}{\bar{B}K^*}  =&\, \frac{\eta^{B\to K^*}}{16 \pi^2} \int\limits_{(M_B + M_{K^*})^2}^\infty ds \frac{ \lamkin^{3/2}(s)}{s^2 (s - Q^2)^2}
        \, |A_0^{B\to K^*}(s)|^2 , \\
    \schi{J=1}{A}{\bar{B}K^*}  =&\, \frac{\eta^{B\to K^*}}{24 \pi^2} \int\limits_{(M_B + M_{K^*})^2}^\infty ds \frac{ \lamkin^{1/2}(s)}{s^2 (s - Q^2)^3}
        \Big( 32 \, M_B^2 M_{K^*}^2 \, |A_{12}^{B\to K^*}(s)|^2 \nonumber \\[-5mm]
        & \hspace{6.2cm} +\, (M_B + M_{K^*})^2 \, |A_1^{B\to K^*}(s)|^2 \Big) , \\
    \schi{J=1}{T}{\bar{B}K^*}  =&\, \frac{\eta^{B\to K^*}}{24 \pi^2} \int\limits_{(M_B + M_{K^*})^2}^\infty ds \frac{ \lamkin^{3/2}(s)}{s (s - Q^2)^4}
        \, |T_1^{B\to K^*}(s)|^2 , \\
    \schi{J=1}{AT}{\bar{B}K^*} =&\,\frac{\eta^{B\to K^*}}{24 \pi^2} \int\limits_{(M_B + M_{K^*})^2}^\infty ds \frac{ \lamkin^{1/2}(s)}{ (s - Q^2)^4}
        \bigg( \frac{ 8 M_B^2 M_{K^*}^2}{(M_B + M_{K^*})^2} \, |T_{23}^{B\to K^*}(s)|^2  \nonumber \\[-5mm]
        & \hspace{6cm} +\, \left(M_B^2 - M_{K^*}^2\right)^2 \, |T_2^{B\to K^*}(s)|^2 \bigg),
    \label{eq:chi_para_AT}
\end{align}
where $\lamkin$ was defined below \eqref{eq:th:T23}.
The contributions arising from $H=\bar{B}_s \phi$ can be obtained
from those arising from $H=\bar{B}K^*$ with obvious replacements.
We assume that isospin symmetry holds for the transition FFs discussed here.
The isospin factor $\eta^{B\to M}$ is equal to 2 for $\bar{B} \to \bar{K}$ and $\bar{B} \to \bar{K}^*$ transitions, while is equal to 1 for $\bar{B}_s  \to \phi$.
\\

The dispersive bound is now obtained by equating the OPE result and the unitarity relation in \autoref{eq:ImPi}:
\begin{align}
    \label{eq:th:bounds:dispbou}
    \schi{J}{\Gamma}{\mathrm{OPE}}
    =
    \schi{J}{\Gamma}{\mathrm{1pt}}
    +
    \schi{J}{\Gamma}{\bar{B}K}
    +
    \schi{J}{\Gamma}{\bar{B}K^*}
    +
    \schi{J}{\Gamma}{\bar{B}_s \phi}
    + \dots \,.
\end{align}
Here, the ellipsis denotes the contribution of further states with the right quantum numbers, such as $\Lambda_b \bar\Lambda$ or $\bar{B} K\pi\pi$.
Each individual hadronic contribution to $\schii{J}{\Gamma}$ is a manifestly positive quantity since it can be expressed as
the modulus squared of a hadronic matrix element. Hence, Eq.~(2.38) can be converted from equality
to inequality simply by dropping unknown but positive exclusive contributions.
Here, we drop the terms represented by the ellipsis.
Using our knowledge of $\schi{J}{\Gamma}{\mathrm{OPE}}$,
\autoref{eq:th:bounds:dispbou} now provides an upper bound on some positive definite
integral involving at least one of $\bar{B} \to \bar{M}$ FF.
This inequality is commonly called the \emph{dispersive (or unitarity) bound}.

\subsection{Dispersively Bounded Parametrization}
\label{sec:th:param}

The dispersive bound of \autoref{eq:th:bounds:dispbou} is not written in a form
that allows us to use it easily in phenomenological applications.
To express it in a more convenient form, the dispersively bounded $z$-series parametrization
has been developed~\cite{Boyd:1997kz,Caprini:1997mu,Bharucha:2010im,Caprini:2019osi,Gubernari:2020eft}.
In the following, we revisit the parametrization of Ref~\cite{Gubernari:2020eft} and propose minor modifications
that improve its usefulness and accuracy.
To this end, we first briefly review the analytic structure of the $\bar{B}\to \bar{M}$ FFs.
As anticipated in \autoref{sec:ff-def}, the FFs arising from a current $J_\Gamma^\mu$
develop a branch cut at the first multi-particle threshold, that is $q^2 \geq s_\Gamma$.
For vector and tensor FFs, this threshold corresponds to the $\bar{B}_s\pi^0$ pair production: $s_V = s_T = (M_{B_s} + M_{\pi^0})^2$.
For axial and axial-tensor FFs, simultaneous conservation of angular momentum and parity
requires at least two pions in the final state, leading to
$s_A = s_{AT} = (M_{B_s} + 2 M_{\pi^0})^2$.

By applying the mapping \autoref{eq:zmap} to the dispersive representation of the two-particle contributions in \Eqs{eq:chi_time_V}{eq:chi_para_AT}, we can rewrite \autoref{eq:th:bounds:dispbou} as
\begin{align}
    \label{eq:th:bounds:dispbou3}
    1
    >
    \sum
    \int\limits_{-\alpha_\Gamma^{B\to M}}^{+\alpha_\Gamma^{B\to M}} \!\!\!\!d\theta \left|\hFM{\Gamma,\lambda}(e^{i\theta})\right|^2
    \, ,
\end{align}
where the sum runs over the transitions $B\to M$ considered in this work and the indices $\lambda$ of the
contributing form factors.
We further abbreviate
\begin{equation}
    \alpha_\Gamma^{B\to M} \equiv \arg z((M_B + M_M)^2; s_\Gamma, s_0)\,.
\end{equation}
In writing \autoref{eq:th:bounds:dispbou3}, we introduce the functions $\hFM{\Gamma,\lambda}$, which are related to the FFs $\FM{\lambda}$ as introduced in \autoref{sec:ff-def}.
Their definition reads
\begin{align}
    \label{eq:hFM}
    \hFM{\Gamma,\lambda}(z)
    =
    \P_\F(z)
    \phi_\F(z)
    \FM{\Gamma,\lambda}(z)
    \,,
\end{align}
where $\P_\F$ is the FF's Blaschke factor, and $\phi_\F$ is the FF's outer function.
The Blaschke factors $\P_\F$ are needed to cancel any simple poles in $\FM{\Gamma,\lambda}$ within
the open unit disk in the $z$ plane. These poles are due to the bound states $R_\F$ discussed
in \autoref{sec:th:bounds}, listed in \autoref{tab:resonances} alongside the corresponding FFs.
We define the Blaschke factors as 
\begin{equation}
    \P_\F(z) \equiv z(s(z); s_\Gamma, M_{R_\F}^2) \,.
\end{equation}
The expressions~\eqref{eq:chi_time_V}-\eqref{eq:chi_para_AT}, once normalised by their respective $\schi{J}{\Gamma}{\mathrm{OPE}}$, fix the norms of the outer functions $\phi_\F$ on the unit circle
\begin{equation}
    |\phi_\F(z)|^2 = 
        \left|\frac{dz}{d\theta} \frac{ds}{dz}\right| 
        \frac{\N_\F \, \eta^{B\to M}}{32 \, \pi^2 \schi{J}{\Gamma}{\mathrm{OPE}}} \,
        \frac{\lamkin^{m/2}}{s^p \, (s - Q^2)^{n+1}} \,
        \qquad
        \text{for } s = s(z = e^{i\theta}).
\end{equation}
Here the parameters $\{\N_\F, p, n, m\}$ are FF-specific and listed in \autoref{tab:outer_parameters}.
Given that the modulus squared of an outer function is fixed only for $|z| = 1$,
there is some leeway in how to continue the outer functions into the open unit disk.
One requirement is that they be free of (kinematical) singularities in
the open unit disk. Another one is to choose their overall phase to be real-valued on the real $z$ axis,
reflecting the fact that their corresponding FFs only develop a branch cut
at $s_\Gamma$ and their global phase can be chosen to be zero.
Setting $Q^2 = 0$, one obtains (see e.g. Ref.~\cite{Boyd:1997kz,Bharucha:2010im,Caprini:2019osi})
\begin{equation}
    \label{eq:outer_function}
    \phi_\F(z)  = \sqrt{\frac{\N_\F \, \eta^{B\to M}}{32 \, \pi^2 \schi{J}{\Gamma}{\mathrm{OPE}}}}
        \left( \frac{\lamkin}{-z(s,s_-)} \right)^{\frac{m}{4}}
        \left( \frac{-z(s,0)}{s} \right)^{\frac{n + p + 1}{2}}
        \sqrt{\frac{4 (1 + z) (s_\Gamma - s_0)}{(z - 1)^3}}
        \,.
\end{equation}

\begin{table}[t]
    \centering
    \setlength{\tabcolsep}{10pt}
    \begin{tabular}{@{}l c c c c@{}}
        \toprule
        Form factor       & $\N_\F$                     & $p$ & $n$ & $m$ \\
        \midrule
        $f_0^{B\to P}$    & $ s_+ s_-$                  &   2 &   1 &  1  \\
        $f_+^{B\to P}$    & $1$                         &   2 &   2 &  3  \\
        $f_T^{B\to P}$    & $1 / s_+$                   &   0 &   3 &  3  \\
        \midrule
        $V^{B\to V}$      & $2 /  s_+$                  &   1 &   2 &  3  \\
        $A_0^{B\to V}$    & $1$                         &   2 &   1 &  3  \\
        $A_1^{B\to V}$    & $2 \, s_+$                  &   1 &   2 &  1  \\
        $A_{12}^{B\to V}$ & $64 \, M_B^2 M_M^2$         &   2 &   2 &  1  \\
        $T_1^{B\to V}$    & $2$                         &   1 &   3 &  3  \\
        $T_2^{B\to V}$    & $2 \, s_+ s_-$              &   1 &   3 &  1  \\
        $T_{23}^{B\to V}$ & $16 \, M_B^2 M_M^2 / s_+$   &   0 &   3 &  1  \\
        \bottomrule
    \end{tabular}
    \caption{
        \label{tab:outer_parameters}
        Parameters of the outer functions of the $\bar{B}\to \bar{P}$ and $\bar{B}\to \bar{V}$ FFs,
        with $s_\pm = (M_B \pm M_M)^2$.
    }
\end{table}

Since $\hFM{\Gamma,\lambda}(z)$ is analytic on the open unit disk, it can be expanded in a Maclaurin series,
\begin{equation}
    \label{eq:th:param:BGL-like}
    \hFM{\Gamma,\lambda} = \sum_{n=0}^\infty b_n^{\F} z^n\,.
\end{equation}
The BGL parametrization only works in the special case where $s_\Gamma = s_+ \equiv (M_B + M_M)^2$, such that $\alpha_\Gamma^{B\to M} = \pi$.
This implies that
\begin{equation}
    \int\limits_{-\pi}^{+\pi} d\theta \, \left|\hFM{\Gamma,\lambda}(e^{i\theta})\right|^2 
        = \sum_{n=0}^\infty \left| b_n^{\F}\right|^2
        < 2\pi\,,
\end{equation}
where the equality follows from the fact that the $z$ monomials are orthogonal on the unit circle.
Since here we deal with the case where $s_\Gamma < s_+$, we follow Ref.~\cite{Gubernari:2020eft}
and instead of using \autoref{eq:th:param:BGL-like} expand $\hFM{\Gamma,\lambda}(z)$ in a series of polynomials $p_n(z)$ that are
orthonormal on a symmetric arc on the unit circle
$\lbrace e^{i\theta} : -\alpha_\Gamma^{B\to M} < \theta <  \alpha_\Gamma^{B\to M}\rbrace$:
\begin{equation}
    \label{eq:hFexp}
    \hFM{\Gamma,\lambda}(z) = 
    \sum_{n \geq 0} a^\F_n \, p_n^\F(z)
\end{equation}
with
\begin{equation}
    \label{eq:th:scalar_product}
    p_n^\F(z) \equiv p_n^\F(z, \alpha_\Gamma^{B\to M}) \,,
    \qquad
    \int\limits_{-\alpha_\Gamma^{B\to M}}^{+\alpha_\Gamma^{B\to M}} \!\!\!\!d\theta \,
    p_m^\F (e^{i\theta}) p_n^\F(e^{-i\theta})
    =
    \delta_{mn}
    \,,
\end{equation}
and hence \autoref{eq:hFM} can be written as
\begin{equation}
    \label{eq:z_exp}
    \FM{\Gamma,\lambda}(z) = \frac{1}{\P_\F(z)\phi_\F(z)} \sum_{n \geq 0} a^\F_n \, p_n(z, \alpha_\Gamma^{B\to M})\,.
\end{equation}
Here it suffices to say that $p_n^\F$ is a polynomial of degree $n$.
Some mathematical properties of these objects as well as a discussion on the convergence of the series $a^\F_n$ are relegated to \autoref{app:orthonormal_polynomials}.
The advantage of using the expansion \autoref{eq:hFexp} is that the dispersive bound \autoref{eq:th:bounds:dispbou3} --- including the one-particle contribution --- can now be written in the simple form
\begin{align}
    \label{eq:th:bounds:dispbou4}
    \frac{\schi{J}{\Gamma}{\mathrm{1pt}}}{\schi{J}{\Gamma}{\mathrm{OPE}}}
    +
    \sum_{\F}
    \sum_{n \geq 0}
    \left|a^\F_n\right|^2
    \, < 1,
\end{align}
where the first sum runs over all the FFs $\FM{\Gamma,\lambda}$.

To summarise, using unitarity and analyticity one obtains the dispersive bound \autoref{eq:th:bounds:dispbou}.
%
Applying a conformal mapping, we recast the bound~\autoref{eq:th:bounds:dispbou} in the more convenient form of \autoref{eq:th:bounds:dispbou4}. 
In doing this, we took into account the analytic properties of the FFs $\FM{\Gamma,\lambda}$ and in particular the branch cuts that appear for $q^2<(M_B + M_M)^2$ due to on-shell $\bar{B}_s  \pi^0 (\pi^0)$ states.
These branch cuts have always been neglected for $B$-meson decays,
which introduces a hard-to-quantify systematic uncertainty. 
Here, we take them into account following the procedure proposed in Ref.~\cite{Gubernari:2020eft} for the non-local FFs in $\bar{B}\to \bar{M} \ell^+\ell^-$ decays.
The dispersive bound~\autoref{eq:th:bounds:dispbou4} is an extremely powerful and model-independent constraint on the coefficients of the expansion~\autoref{eq:hFexp}.
In practice, it allows us to control the systematic uncertainties due to the truncation of same expansion.

\section{Analysis and results}
\label{sec:analysis}

\subsection{Analysis setup}
\label{sec:analysis:setup}

\newcommand{\vth}{\ensuremath{a}}
\newcommand{\vecth}{\ensuremath{\vec{\vth}}}

We now study the available theoretical data on the FFs within a Bayesian analysis. Its central element, the posterior PDF $P(\vecth\,|\, D)$,
is a function of the data $D$ and our parameters $\vecth \equiv (a^\F_0, \dots, a^\F_N, a_0^{\F'}, \dots, a_N^{\F'}, \dots)$.
We use the same value of $N$ as the order of truncation of the series \autoref{eq:z_exp} for each FF. The posterior definition
\begin{equation}
    P(\vecth\,|\, D) = \frac{P(D \,|\, \vecth) P_0(\vecth)}{Z(D)}
\end{equation}
involves our choice of the prior PDF $P_0(\vecth)$, the likelihood $P(D\,|\,\vecth)$, and the data-dependent normalisation (also called the evidence)
$Z(D)$.\\

We use different choices of fit model and prior. We label all our fits models
by $N \in \{2, 3, 4\}$, the truncation order applying to all FFs.
The choice of the prior PDF is motivated by the weakest possible form of the dispersive bound, i.e., $|\vth_k| \leq 1$ for all $k$.
This prior factorises to
\begin{equation}
    P_0(\vecth) = \prod_{k=1}^{K=\dim \vecth} \mathcal{U}_{-1,+1}(\vth_k)\,,
\end{equation}
where $\mathcal{U}_{-1,+1}$ is the uniform PDF on the interval $[-1, +1]$.
Since we have 17 independent FFs and 9 end-point relations, the number of independent parameters is $K = 17 (N+1) - 9$.\\

We face two types of likelihoods in this analysis:
linear constraints on the FF parameters provided by LQCD and LCSR determinations;
and 
quadratic constraints on the FF parameters provided by the dispersive bounds \autoref{eq:th:bounds:dispbou4}.
The linear multivariate Gaussian likelihoods arising from a single hadronic transition are labelled appropriately, i.e., either $B \to K$, $B \to K^*$, or $B_s \to \phi$. The data underlying each likelihood are summarised as follows:
\begin{description}
    \item[$\boldsymbol{B \to K}$]
    We use three LQCD analyses of the three $\bar{B}\to \bar{K}$ FFs.
    The underlying LQCD analyses have been performed by the Fermilab Lattice and MILC collaborations (FNAL + MILC) \cite{Bailey:2015dka} and by the HPQCD collaboration ~\cite{Bouchard:2013eph, Parrott:2022rgu}.
    Although an average of the FNAL+MILC and the 2013 HPQCD results is available from the Flavour Lattice
    Averaging Group (FLAG)~\cite{Aoki:2021kgd}, we choose to use the individual results rather than their average
    here. This enables us to diagnose which of the individual analyses are in mutual agreement.
    As indicated in the FLAG report, the FNAL+MILC and the 2013 HPQCD analyses rely --- at least partially ---
    on the same underlying gauge ensembles. However, their overall uncertainties are dominated by systematic
    sources, rather than the purely statistical uncertainty. Hence, we combine the results of
    both analyses under the assumption that they are uncorrelated, just as has been done to obtain the FLAG average~\cite{Aoki:2021kgd}.
    The third analysis, HPQCD 2022~\cite{Parrott:2022rgu}, uses $N=2+1+1$ gauge ensembles that are
    statistically independent of the previous analyses.
    The results of the three analyses are provided as multivariate Gaussian distributions in the parameters of a BCL expansion~\cite{Bourrely:2008za}.
    We use the parametrized results of the various LQCD analyses to produce synthetic data points across the three FFs,
    making sure that the number of data points matches the number of degrees of freedom in the parametrizations.
    For the FNAL+MILC and HPQCD 2013 analyses, we generate points close to the $q^2$ values of the lattice ensembles, $q^2 = \{17, 20, 23\} \, \mathrm{GeV}^2$.
    The HPQCD 2022 results provides for the first time access to an exclusive $\bar{B} \to \bar{K}$ FF at $q^2 = 0$.
    Consequently, we spread the points over the entire physical range, $q^2 = \{0, 12, 22.9\} \, \mathrm{GeV}^2$.
    \\
    
    The $\bar{B} \to \bar{K}$ FFs are also available from LCSR analyses with kaon distribution amplitudes. The most recent results obtained in this framework are available from Ref.~\cite{Khodjamirian:2017fxg}.
    Compared to the LQCD results, these estimates feature large parametric uncertainties and
    hardly-quantifiable systematic uncertainties. As a consequence, we do not use these estimates
    to obtain any of our numerical results and only use them for illustrative purposes and 
    cross checks.

    \item[$\boldsymbol{B \to K^*}$]
    The full set of FFs is available from a LQCD analysis and a subsequent addendum~\cite{Horgan:2013hoa,Horgan:2015vla}. The authors of that analysis
    have provided us with twelve synthetic data points per FF, corresponding to
    the extrapolation of their LQCD ensembles.
    Correlations between the FFs arising from (axial)vector
    and (axial)/tensor currents are not presently available and can therefore not be accounted
    for in our analysis.\\
    
    Beside from this LQCD analysis, numerical results for the full set of FFs are also
    available from LCSR with $B$-meson distribution amplitudes~\cite{Gubernari:2018wyi}.
    The authors of this calculation provide five synthetic data points at $q^2 = \{-15, -10, -5, 0, 5\} ~\mathrm{GeV}^2$ for each FF,
    and the full correlation matrix across FFs and data points are available.
    Following the argument in our previous paper~\cite{Gubernari:2022hxn}, we do not use LCSR with light-meson distribution amplitude for FFs with vector final states. 
   
    \item[$\boldsymbol{B_s \to \phi}$]
    The inputs are similar to the $\bar{B} \to \bar{K}^*$ ones.
    Refs.~\cite{Horgan:2013hoa,Horgan:2015vla} provide the mean to create two synthetic data points per FF, reflecting the smaller number of LQCD ensembles for this transition.
    Ref.~\cite{Gubernari:2020eft} provides LCSR estimates of these FFs using a previous calculation published in Ref.~\cite{Gubernari:2018wyi}.
    Five synthetic data points including their correlations are available at $q^2 = \{-15, -10, -5, 0, 5\} ~\mathrm{GeV}^2$ for each FF.
\end{description}

The dispersive bounds discussed in \autoref{sec:th:bounds} are labelled by their respective Dirac structure~$\Gamma$ and spin $J$.
The penalty for violating the dispersive bounds is implemented following Ref.~\cite{Bordone:2019vic}:
\begin{equation}
     \label{eq:penalty_likelihood}
    -2\ln P(\text{dispersive bound} \, | \,  r_{\Gamma,J}) 
        = \begin{cases}
            0    &   \text{if }  r_{\Gamma,J} < 1, \\
            \dfrac{\left( r_{\Gamma,J} - 1\right)^2}{\sigma^2}
                 & \text{otherwise.}
         \end{cases}\,
\end{equation}
Here $\sigma$ reflects the relative uncertainty on the computation of $\schi{J}{\Gamma}{\mathrm{OPE}}$, which is estimated to be $10\%$~\cite{Bharucha:2010im},
and $r_{\Gamma,J}$ is defined as
\begin{equation}
   r_{\Gamma,J} \equiv
    r_{\Gamma,J}^{\mathrm{1pt}}
    + r_{\Gamma,J}^{B\to K}
    + r_{\Gamma,J}^{B\to K^*}
    + r_{\Gamma,J}^{B_s\to \phi}
    =
    \frac{\schi{J}{\Gamma}{\mathrm{1pt}}}{\schi{J}{\Gamma}{\mathrm{OPE}}}
    +
    \sum_{\F}
    \sum_{n = 0}^N
    \left|a^\F_n\right|^2.
\end{equation}
The quantity $r_{\Gamma,J}$ is the statistical estimator for the saturation of the bound with spin $J$ for the current $J_\Gamma^\mu$
defined in \autoref{eq:th:ff-def:JGamma}.

Imposing the dispersive bounds as expressed in \autoref{eq:th:bounds:dispbou} makes a combined analysis of the FFs
worthwhile but also challenging.
It is worthwhile since in the presence of the dispersive bounds some of the parametrization uncertainties are
shared among the fitted transitions.
It is challenging due the large number of fit parameters $K$, e.g., $K=59$ in our nominal analysis with $N=3$.
%
Since a joint fit to the form factors in all three transitions involves only correlations
due to the dispersive bound, we can simplify the analysis somewhat.
Separating $\bar{B}\to \bar{K}$, $\bar{B} \to \bar{K}^*$ and $\bar{B}_s \to \phi$ transitions
into individual fits, we can achieve the results of the joint fit by re-weighting their individual parameter samples
with their dispersive bounds' likelihood.
Taking $\bar{B} \to \bar{K}^*$ as an example, the current-specific sample weight is computed as
\begin{equation}
    w(r_{\Gamma,J}^{B\to K^*})
        = \iint dr dr' \, p_{\Gamma,J}^{B_s\to \phi}(r) \, p_{\Gamma,J}^{B\to K}(r') \,
            \times P\left(\text{dispersive bound for $\Gamma, J$} \, |\, r_{\Gamma,J}^\mathrm{1pt} +  r_{\Gamma,J}^{B\to K^*} + r + r'\right),
\end{equation}
where $r_{\Gamma,J}^\mathrm{1pt}$ and $r_{\Gamma,J}^{B\to K^*}$  are the saturations of the corresponding bound due to the one particle and the $\bar{B} \to \bar{K}^*$ process respectively,
and $p_{\Gamma,J}^{B_s\to \phi}$, $p_{\Gamma,J}^{B \to K}$ are the PDFs of the saturation due to the \mbox{$\bar{B}_s \to\phi$}
and the \mbox{$\bar{B}\to K$} processes, respectively.
The global event weight is obtained by multiplying these statistically independent weights,
\begin{equation}
    w^{B\to K^*} = \prod_{\Gamma,J} w(r_{\Gamma,J}^{B\to K^*})\,.
\end{equation}
This procedure can be extended to account for other transitions, e.g., baryonic transitions.\\
It permits us to split the production of Monte Carlo samples for our joint analysis into three individual samplings, one per transition,
which facilitates the overall analysis significantly.\\

We carry out the fitting and sampling parts of our analysis using the open-source \EOS software~\cite{EOSAuthors:2021xpv} version v1.0.7~\cite{EOS:v1.0.7}.
For each value of the truncation order $N$, we determine the best-fit point and overall goodness-of-fit diagnostics.
Our a-priori $p$-value threshold for an acceptable fit is $3\%$.
We draw all posterior samples using the nested sampling algorithm~\cite{Higson:2018} as implemented in the \texttt{dynesty} software~\cite{Speagle:2020,dynesty:v2.0.3}.
The results are investigated in form of posterior-predictive distributions for a variety of pseudo observables,
including all of the FFs and the relative saturation of the various dispersive bounds.

\subsection{Numerical results}

Our analysis is repeated for three values of the truncation order $N \in \{2, 3, 4\}$, as anticipated in the previous section.
In agreement with the literature~\cite{Bharucha:2010im,Bouchard:2013eph,Horgan:2013hoa,Horgan:2015vla,Bailey:2015dka,Bharucha:2015bzk,Gubernari:2018wyi,Parrott:2022rgu},
we already find a good fit to \emph{all} FFs at $N = 2$.
The substantial uncertainties attached to both LQCD and LCSR estimates of the FFs lead to local $p$-values close to $1$ for both $\bar{B} \to \bar{K}^*$ and $\bar{B}_s \to \phi$ transitions; we find all local $p$-values to be in excess of $77\%$.\footnote{%
    Here the local $p$-value refers to the computation of the $p$-value of a single likelihood given its effective degrees of freedom,
    \ie, the number of observations minus the number of fit parameters specific to this likelihood.
    Neither the local $p$-values nor the global $p$-value account for the non-Gaussian effect of the dispersive bounds,
    which we do not include in our discussion of the goodness of fit whatsoever.
}
We summarise the goodness-of-fit diagnostics for $N = 2$ and $N = 3$ in \autoref{tab:numerics:gof}
and we consider the $N = 3$ analysis to yield our nominal results.
We find that higher truncation orders do not have an adverse impact on the overall goodness of fit for nearly all individual likelihoods.
However, in the case of $\bar{B}_s\to \phi$ the $\chi^2$ value is so low that increasing the truncation order beyond
$N=3$ can only lower the local $p$-value.

\begin{figure}[t]
    \centering
    \includegraphics[width=0.7\textwidth]{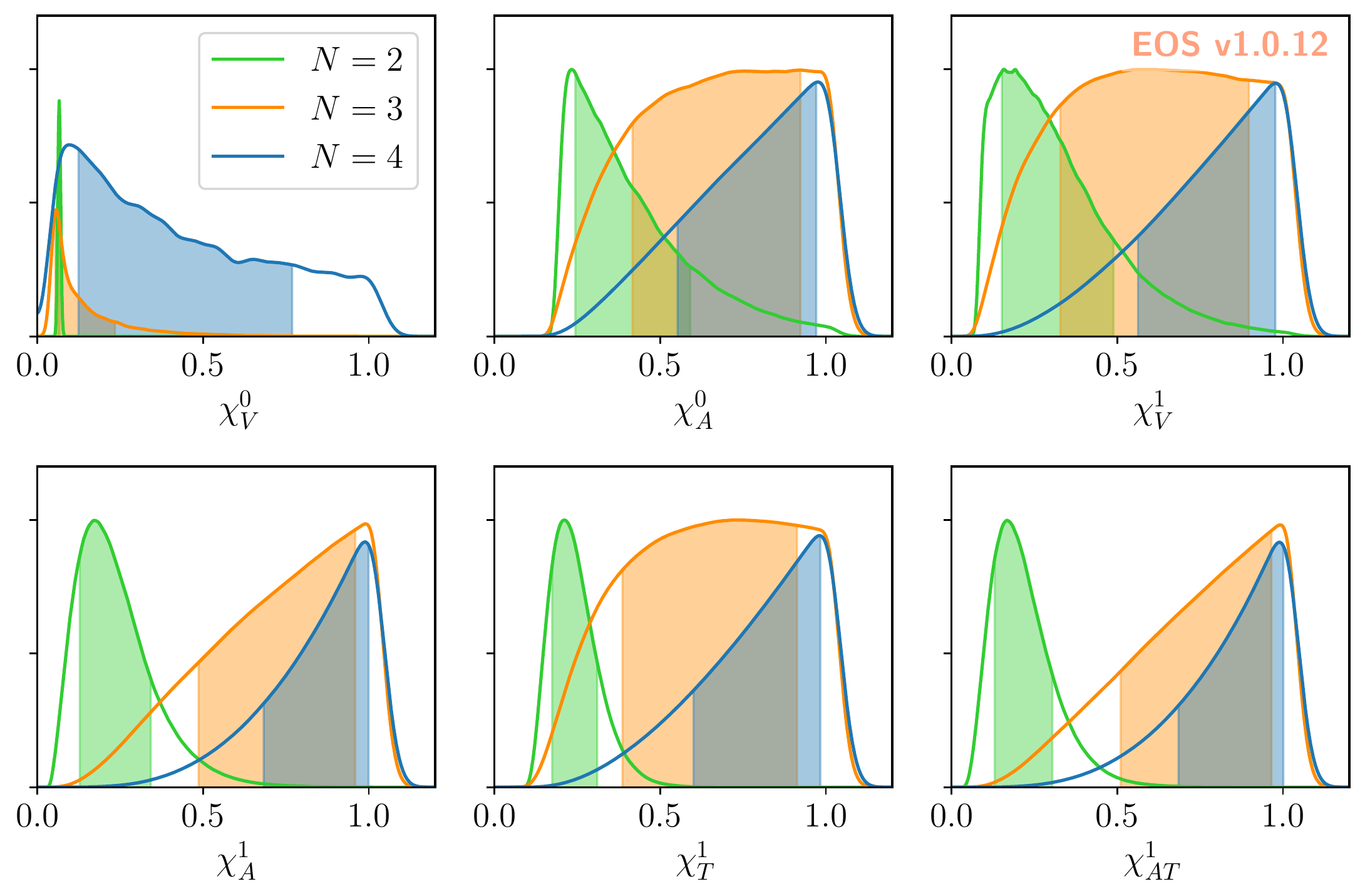}
    \caption{%
        Saturations of the dispersive bounds due to one- and two-particle contributions.
        The lines represent Gaussian-smoothed distributions of the saturations in our samples that we combined following the discussion in \autoref{sec:analysis:setup}.
        The shaded area comprise the $68\%$ probability interval for each scenario.
        The distributions are scaled with arbitrary factors for readability.
    }
    \label{fig:results:saturation}
\end{figure}

We illustrate the results of our analyses at the hand of a representative subset of FFs in \autoref{fig:local_ffs}.
Plots for the full set of FFs are contained in the supplementary material~\cite{EOS-DATA-2023-02}, alongside the
\EOS analysis file that has been used to obtain all numerical results shown in this section.
Using \autoref{eq:th:bounds:dispbou4}, we further compute the saturations of the full set of dispersive bounds
and present them in \autoref{tab:numerics:saturations} and \autoref{fig:results:saturation}. We use this
information later on to determine if the parametric uncertainties account for the truncation error.\\

\begin{table}[t]
    \centering
    \setlength{\tabcolsep}{10pt}
    \begin{tabular}{@{}l ccc ccc@{}}
        \toprule
                                & \multicolumn{6}{c}{Goodness of fit}
        \\
                                & \multicolumn{3}{c}{$N=2$}             & \multicolumn{3}{c}{$N=3$}
        \\
        Transition              & $\chi^2$  & d.o.f.  & $p$-value [\%]  & $\chi^2$  & d.o.f.  & $p$-value [\%]
        \\
        \midrule
        $\bar{B}\to \bar{K}$    & $13.32$   & $18$    & $77.25$         &  $7.42$   & $15$    & $94.48$
        \\
        $\bar{B}\to \bar{K}^*$  & $50.72$   & $100$   & $100$           & $45.77$   & $93$    & $100$
        \\
        $\bar{B}_s\to \phi$     & $1.19$    & $30$    & $100$           &  $0.32$   & $23$    & $100$
        \\
        \bottomrule
    \end{tabular}
    \caption{%
        Goodness-of-fit values for the fit models $N=2$ and $N=3$.
    }
    \label{tab:numerics:gof}
\end{table}

\begin{figure}[t]
    \centering
    \includegraphics[width=.32\textwidth]{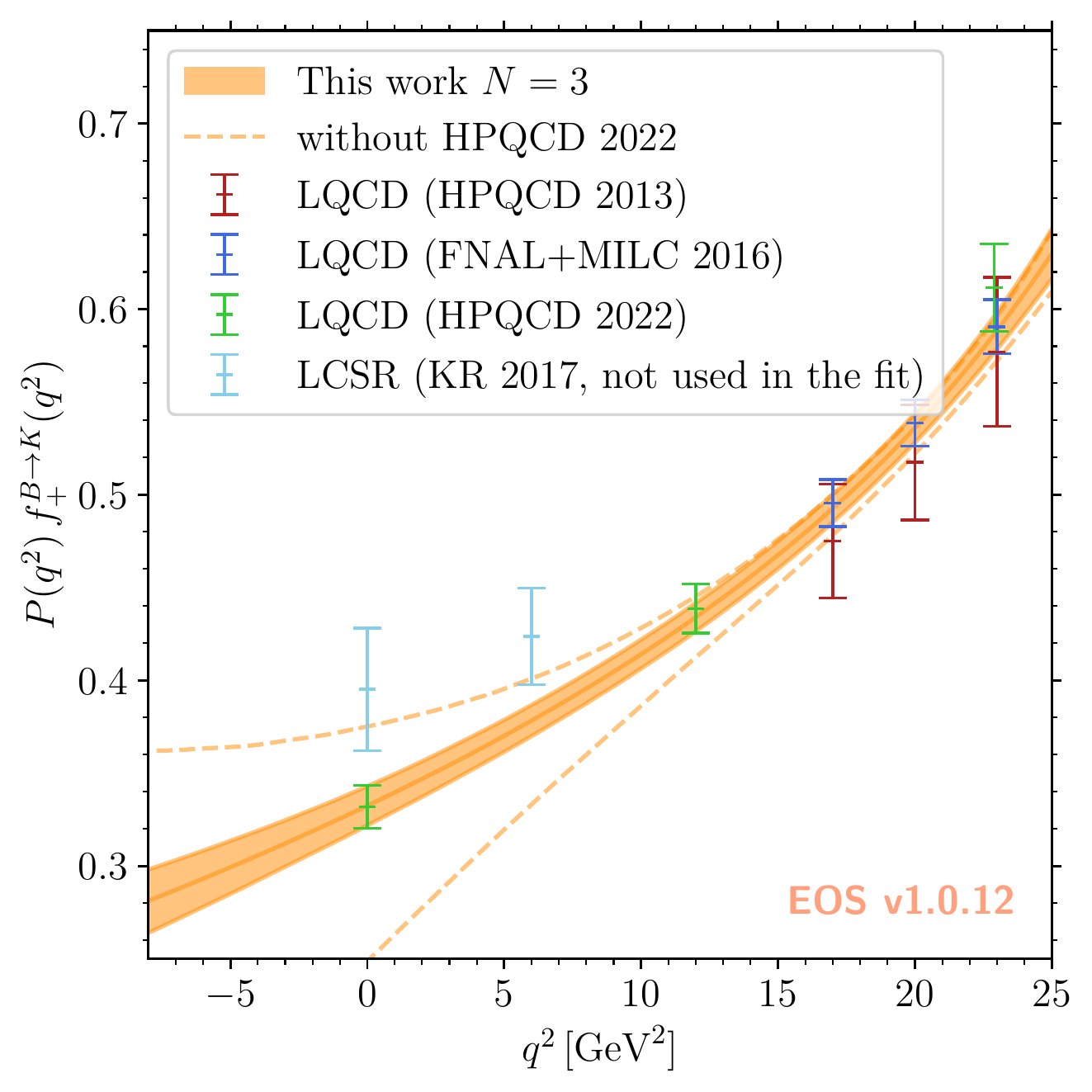}
    \includegraphics[width=.32\textwidth]{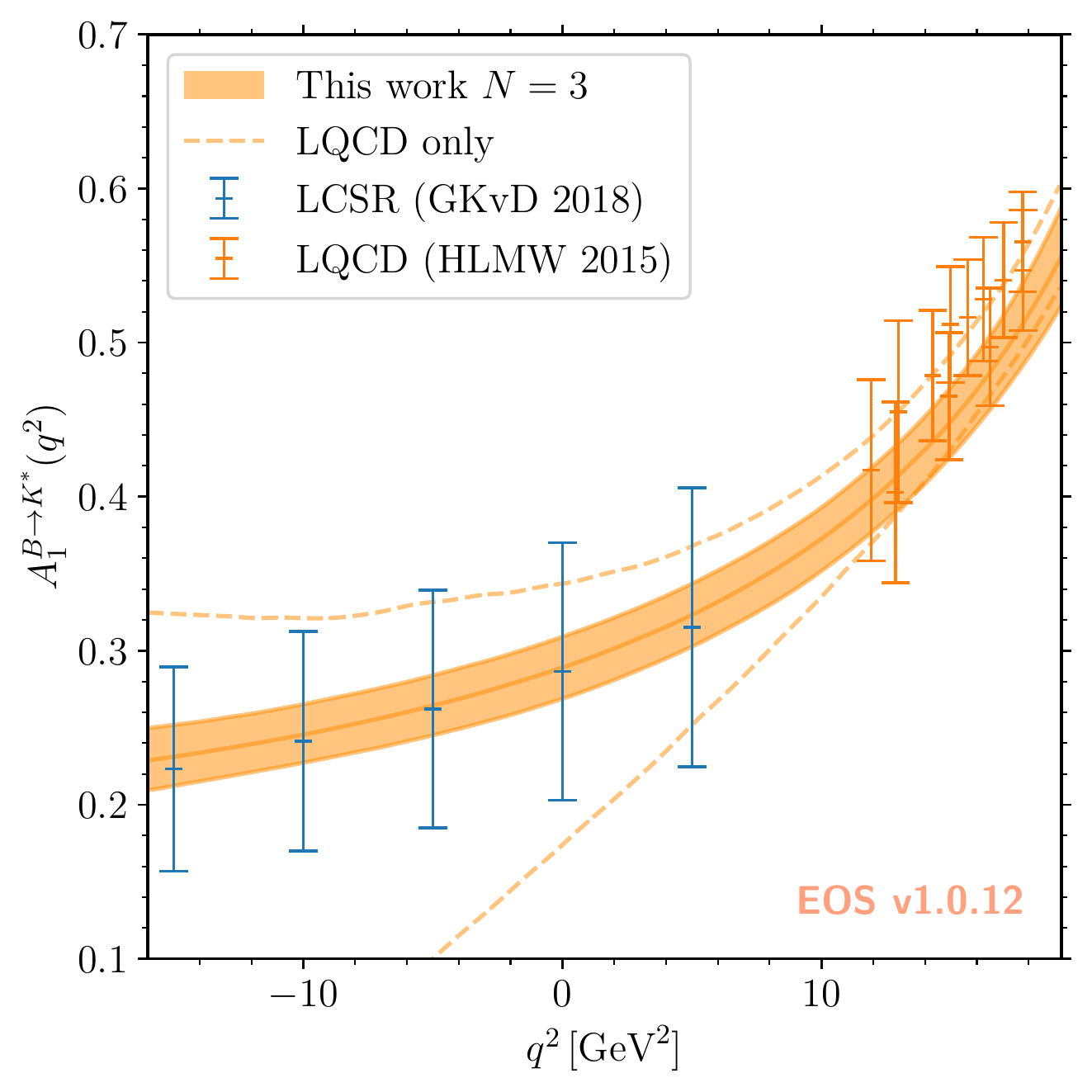}
    \includegraphics[width=.32\textwidth]{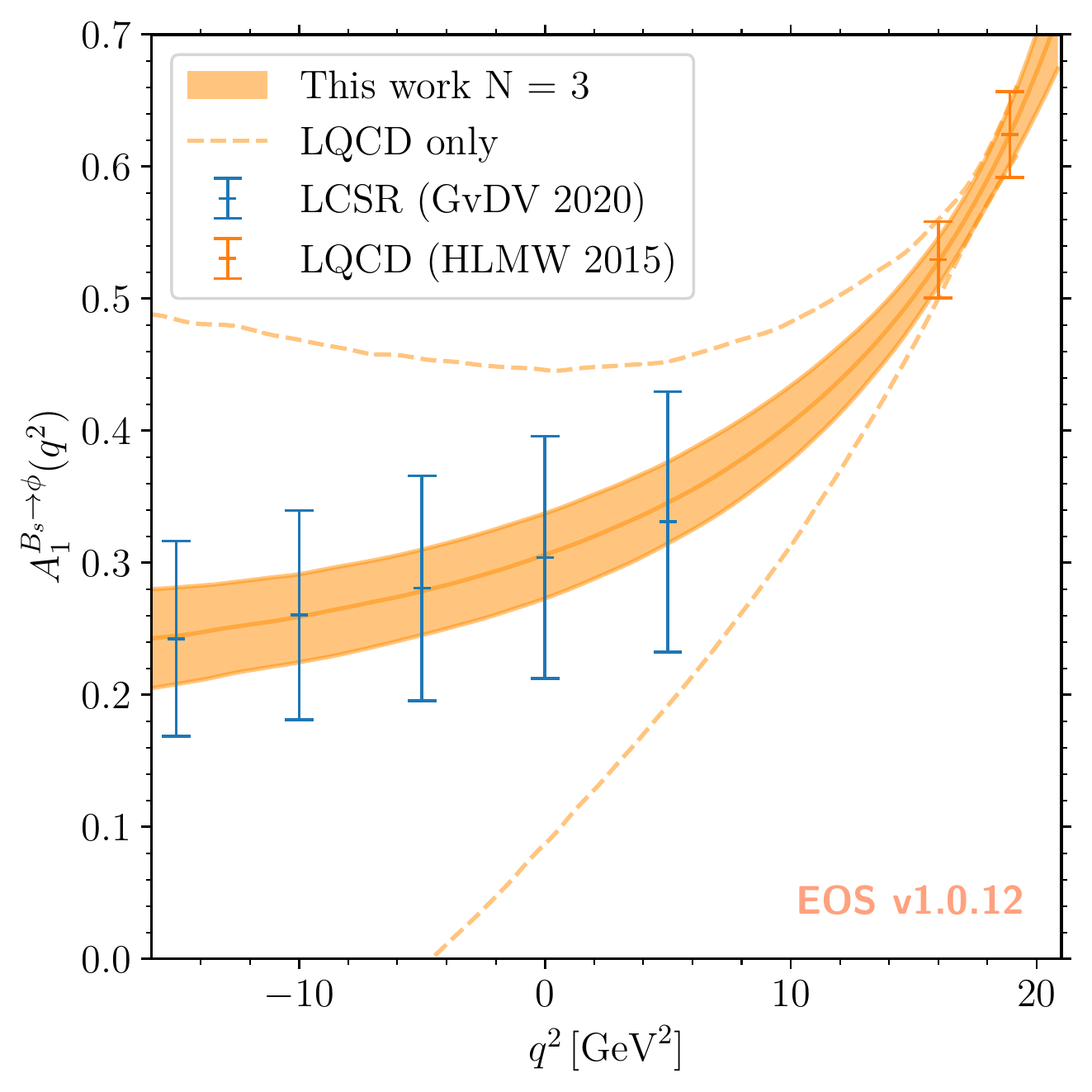}
    \caption{%
        Selection of local $\bar{B}\to \bar{M}$ FFs.
        $f_+^{B\to K}$ is multiplied by $P(q^2) = 1 - q^2 / M_{B_s^*}^2$ for legibility.
        Additional plots, including plots of the other FFs are provided as supplementary material.
    }
    \label{fig:local_ffs}
\end{figure}

We summarise our findings arising from the individual fits as follows:
\begin{description}
    \item[Fit with $\boldsymbol{N=2}$]
    We confirm the observation of Ref.~\cite{Bharucha:2010im} that for this value of $N$, the bounds are not yet saturated and only play a marginal role in the fit.
    We can quantify this finding by determining the saturation at $68\%$ cumulative probability based on the samples' posterior-predictive distributions.
    We find that all the bounds are saturated at less than $35\%$.
    The bound for $(\Gamma, J) = (V, 0)$, which receive two-particle contributions only from $\bar{B}\to \bar{K}$ FFs,
    exhibit a very low saturation below $10\%$.

    \item[Fit with $\boldsymbol{N=3}$]
    As shown in \autoref{tab:numerics:gof}, increasing the truncation order leads to a decrease of both $\chi^2$ and $\chi^2$/d.o.f. across all the transitions.
    Increasing the truncation order mostly impacts the FF values and uncertainties at low and at negative $q^2$, which is expected
    due to the dominance of the LQCD information at high $q^2$ values.
    Since the FF results in the region of low and negative $q^2$ are crucial for the estimation of non-local effects~\cite{Gubernari:2022hxn}, this
    behaviour raises concerns toward the uncertainty estimation for the non-local effects.\\

    We observe that the saturations of the posterior predictive samples now peak at larger values.
    We quantify this change by comparing the saturations at $68\%$ cumulative probability for both $N=2$ and $N=3$ in \autoref{tab:numerics:saturations}.
    Although the bound for $(\Gamma,J) = (V, 0)$ increases slightly in its saturation, we still find it to be saturated
    well below $20\%$ at $68\%$ cumulative probability.
    All other bounds now reach substantial saturations of $69\%$ or larger.
    
    \item[Fit with $\boldsymbol{N\geq 4}$]
    For completeness, we also perform the fit with $N=4$. We find that for all transitions, the minimal $\chi^2$ reaches a plateau, \ie, it does not significantly
    reduce as we increase the truncation order. As indicated above, this inevitably leads to a lowering of its local $p$-value as $N$ increases.
    All bounds now exhibit a saturation of at least $50\%$ at $68\%$ cumulative probability.
\end{description}

We find that the average saturation of all bounds increases with increasing truncation order $N$.
This is expected, since the parametric uncertainties on the posterior predictions of the FFs
start to accurately include the extrapolation uncertainties.
We find that this is achieved for $N \geq 3$.
Our interpretation is supported by
the ``\texttt{comparison}''-type plots contained in our supplementary material~\cite{EOS-DATA-2023-02}; see \autoref{app:supplementary_material}.
Nevertheless, one bound exhibits low saturations below $20\%$ even for $N=3$, which requires further discussion.
The bounds at hand feature $\Gamma=V$ and $J=0$ and receive only contributions from $\bar{B}\to \bar{K}$ FFs
in our analysis.

In a general dispersively-bounded analysis, there is an intrinsic ambiguity in how to interpret the fact that a bound only receives comparatively small increases
to its saturation as we increase the truncation order.\\
On the one hand, the uncertainties attached to the theory estimates of the FFs might prevent our fit model from accurately capturing details on
the shape of the FFs, which can lead to small saturations.\footnote{%
    Recall that in the strict BGL parametrization the fit parameters correspond to a series of derivatives
    of the corresponding FF at the zero-recoil point. Hence, large saturations limit the magnitude
    of these derivatives and vice versa.
}
In this case, the problem can be addressed by increasing the precision of the FFs' theory estimates.\\
On the other hand, our fit model might accurately capture the saturation. Hence the observed low saturation values might be a physical
feature of the responsible FFs.
In this case, one would expect that
including further relevant two-particle transitions and transitions with even higher multiplicities would drive the saturations to higher values.

In our analysis specifically, the data leads us to conclude that the second interpretation is the correct one.
We hence argue that including baryonic transition FFs in the analysis could yield a significant impact
of $\bar{B}\to \bar{K}$ FFs.
As discussed above, the bound for $(\Gamma, J) = (V, 0)$ is only weakly saturated even for $N=3$.
First, we have explicitly checked that this is not leading to underestimated uncertainties in the FF $f_0^{B\to K}$,
which is the only FF that contributes to this bound.
Our check is performed by evaluating this FF for $N=4$ at two phase space points: $q^2 = -5\,\mathrm{GeV}^2$ and $q^2 = +25\,\mathrm{GeV}^2$,
since in this scenario, this bound starts to be saturated.
The largest increase in the uncertainties with respect to the scenario $N=3$ is obtained for $f_+^{B\to K}(-5\,\mathrm{GeV}^2)$ and is of the order of $0.2\%$.
This very small increase in the uncertainties can also be seen in the ``\texttt{comparison}''-type plots contained in our supplementary material~\cite{EOS-DATA-2023-02}.
It is in line with our conclusion that $N \geq 4$ is generally not needed to achieve an accurate estimate of the truncation error for all other FFs,
and therefore further supports our choice of $N = 3$ as the nominal fit.
Next, we compare our results with those of dispersively bounded analyses for baryonic transition FFs as presented in Refs.~\cite{Blake:2022vfl, Amhis:2022vcd}.
We find that in both of these analyses, the saturation of the $\Gamma=V$ bound complements our results.
Our interpretation of these findings is that the $\Gamma=V$ bounds receive substantial contributions to their relative saturation
from the baryonic transitions.
This points to a tangible benefit of analysing mesonic and baryonic transition FFs simultaneously and might lead to
even more precise predictions for the FF $f_0^{B\to K}$.

\begin{table}[t]
   \renewcommand{\arraystretch}{1.2}
   \setlength{\tabcolsep}{20pt}
    \centering
    \begin{tabular}{@{}l c c c c @{}}
        \toprule
        \multirow{2}{*}{$\schii{J}{\Gamma}$} 
                               & \multirow{2}{*}{1-pt saturations [\%]}
                                                   & \multicolumn{3}{c}{2-pt sat.~at 68\% cum.~prob. [\%]} \\
                               &                   & $N = 2$  & $N = 3$ & $N = 4$ \\
        \midrule
        $\schii{J=0}{V}$       &  ---              & $ 6.6$   & $14.6$  & $55.7$ \\
        $\schii{J=0}{A}$       &  $11.8 \pm 0.6$   & $34.4$   & $69.7$  & $78.3$ \\
        $\schii{J=1}{V}$       &  $ 2.8 \pm 0.2$   & $34.4$   & $74.1$  & $88.1$ \\
        $\schii{J=1}{A}$       &  ---              & $27.0$   & $87.5$  & $95.1$ \\
        $\schii{J=1}{T}$       &  $ 4.7 \pm 0.9$   & $22.2$   & $75.1$  & $87.5$ \\
        $\schii{J=1}{AT}$      &  ---              & $24.6$   & $88.5$  & $95.3$ \\
        \bottomrule
    \end{tabular}
    \renewcommand{\arraystretch}{1.0}
    \caption{%
        Relative saturations of the dispersive bounds due to the one-particle and two-particle contributions.
        Two-particle contributions are shown for the different truncation orders and
        the table shows the saturation at $68\%$ cumulative probability.
    }
    \label{tab:numerics:saturations}
\end{table}

\section{Conclusions}
\label{sec:conclusions}

We perform a dispersively bounded analysis by fitting a recently proposed parametrization to existing theoretical data on $\bar{B}\to \bar{K}^{(*)}$ and $\bar{B}_s\to\phi$ local form factors (FFs). 
This data is available from lattice QCD analyses and light-cone sum rule calculations.
The latter data is used to anchor the respective FFs at low $q^2$ whenever lattice QCD results for this region are unavailable. 
The proposed parametrization is needed in our analysis due to the appearance of below-threshold branch cuts in the FFs, generated by $\bar{B}_s \pi^0$ and  $\bar{B}_s \pi^0\pi^0$ states.
These branch cuts have not been considered in previous analyses, giving rise to unwanted and hard-to-quantify systematic uncertainties.

We obtain numerical results for the fitted parameters from a comprehensive Bayesian analysis of the three transitions $\bar{B}\to \bar{K}^{(*)}$ and $\bar{B}_s\to\phi$.
We provide the $N=2$ results in a machine-readable form as supplementary material~\cite{EOS-DATA-2023-02}.
Our results include the central values, standard deviations, and correlations of both the coefficients obtained in our framework and the coefficients of the BSZ parametrization.
The parameters within both parametrizations follow multivariate Gaussian distributions to a good approximation, with relative perplexities in excess of $95\%$.
At the current level of precision on the estimation of the FFs describing $\bar{B} \to \bar{K}^{(*)}$ and $\bar{B}_s \to\phi$ transitions, we find that
at present, a simplified series expansion \emph{\`a la} BSZ provides an accurate estimation of the FF uncertainties at positive $q^2$,
although this may change once more and more precise lattice QCD results become available.
However, at negative $q^2$, the coupled effect of a larger truncation order and the impact of the dispersive bounds are already essential to correctly estimate the extrapolation uncertainties.
This fact is particularly important for the determination of the non-local FFs for these transitions since at low and negative values of $q^2$ they are proportional
to the local FFs at leading power in the light-cone OPE.
Therefore, we conclude that our proposed approach is crucial for this determination.

We emphasise that present and future lattice QCD analyses must fulfil the dispersive constraints.
Hence, we encourage the implementation of our framework to perform \emph{a posteriori} checks.
Nevertheless, we caution that dispersive bounds should not be applied directly or exclusively in theoretical calculations;
doing so would prohibit their use in simultaneous FF fits.\\

Our present analysis is both a benchmark of what is possible within the framework of dispersively bounded FF parametrizations
and a step stone towards more precise predictions of the non-local FFs at negative $q^2$.
A key advantage of this framework is the possibility of including further transitions in the analysis, in particular baryonic ones.
We expect this to lead to a reduction of the uncertainties of all the transitions considered within a global analysis.
Thus, we regard the work presented here as an important step towards a truly global analysis of rare $B$-decay observables
that simultaneously uses dispersive bounds to control all systematic uncertainties, thereby advancing the precision frontier in indirect BSM searches.

\section*{Acknowledgements}

We thank Wolfgang Altmannshofer, Paolo Gambino and Peter Stangl for useful comments on our manuscript.
N.G. thanks Michael Wick for providing the code to evaluate the two-loop corrections to the tensor form factors bound.
M.R.~thanks Guillermo Lopez and Brian Simanek for useful inputs on the asymptotic behaviour of orthonormal polynomials
as well as the IJCLab for its hospitality during the time the paper was written.
The research of N.G. was supported by the Deutsche Forschungsgemeinschaft (DFG, German Research Foundation) under grant 396021762 - TRR 257 “Particle Physics Phenomenology after the Higgs Discovery''.
D.v.D.~acknowledges support by the UK Science and Technology Facilities Council (grant numbers ST/V003941/1 and ST/X003167/1).
J.V.~acknowledges funding from the Spanish MINECO through the ``Ram\'on y Cajal'' program RYC-2017-21870,
the ``Unit of Excellence María de Maeztu 2020-2023'' award to the Institute of Cosmos Sciences (CEX2019-000918-M)
and from the grants PID2019-105614GB-C21 and 2017-SGR-92, 2021-SGR-249 (Generalitat de Catalunya).

\newpage

\appendix

\section{Supplementary material}
\label{app:supplementary_material}

The full set of posterior samples and the \EOS analysis files used to conduct the analysis presented
in this paper are made public as part of our supplementary material~\cite{EOS-DATA-2023-02}.

\subsection{Additional figures}

We provide the following types of figures in addition to the figures contained in this article:
\begin{description}
    \item[\texttt{overview}] These figures show plots of each FFs in relation to the inputs used within their analysis. In the case of $\bar{B}\to \bar{K}$
    FFs, we also illustrate the relation to LCSR results that have not been used in the analysis.
    The names follow the pattern ``\texttt{TRANSITION\_FF.pdf}'', where ``\texttt{TRANSITION}'' can be either
    ``\texttt{BToK}'', ``\texttt{BToKstar}'', or ``\texttt{BsToPhi}``
    and ``\texttt{FF}'' can be either ``\texttt{fp}'', ``\texttt{f0}'', ``\texttt{fT}'', ``\texttt{V}'', ``\texttt{A0}'', ``\texttt{A1}'',
    ``\texttt{T1}'', ``\texttt{T2}'', ``\texttt{T23}``. We do not provide plots for ``\texttt{A2}'' or ``\texttt{A12}'', since
    one of them is redundant and the available constraints are not consistently available for either choice.

    \item[\texttt{comparison}] These figures show plots of the FFs as the envelopes at $68\%$ probability based on their posterior predictive distributions.
    Each figure shows plots of the FF in the dispersively bounded fit for $N=2$ and $N=3$ as well as the results of an additional BSZ fit for $N=2$.
    In each plot, the FF is normalized to its central results of our nominal fit at $N=3$ and in the presence of a dispersive bound.
    Hence, these figures can be used to infer the behaviour of the systematic uncertainty inherent in the extrapolation of the adhoc BSZ parametrization
    and the size of the truncation uncertainty in the dispersively bounded approach.
    The names follow the pattern ``\texttt{TRANSITION\_NFF.pdf}'', where the ``\texttt{N}'' is meant literally;
    see above for the possible values of ``\texttt{TRANSITION}'' and ``\texttt{FF}''.
\end{description}

\subsection{Form factor parameters}

For $N=2$, the effects of the dispersive bounds are too small to distort the distributions of the posterior samples.
The later can therefore be approximated by multivariate Gaussian distributions to a good accuracy, with perplexities of $99.8\%$, $95.1\%$ and $94.2\%$ for $\bar{B} \to \bar{K}$, $\bar{B} \to \bar{K}^*$ and $\bar{B}_s \to\phi$ respectively.
We provide these distributions for the three transitions that are part of our analysis, both for the parameters arising in our parametrization
and also for the BSZ parametrization (i.e., the simplified series expansion of Ref.~\cite{Bharucha:2015bzk}).
These distributions are provided in the ancillary files \texttt{GRvDV-parameters-N2.yaml} and \texttt{BSZ-parameters-N2.yaml} respectively.

Some comments are due on our results in the BSZ parametrization:
\begin{itemize}
    \item We find overall a good fit to the available data even for $N=2$, not unlike the case of our dispersively bounded analysis.
    \item At present, this parametrization accurately covers the uncertainties for the dispersively bounded analysis when we restrict
    the $q^2$ interval to $[0, s_-]$. This is a consequence of the fact that in this region the uncertainties are presently dominated
    by purely parametric sources and the truncation error is not (yet) relevant. It also strengthens BSM analyses that rely on these
    FFs in this phase space and within this parametrization.
    \item At $q^2 < 0$, we start to see that BSZ parametrization does not faithfully capture the uncertainty envelope of the
    dispersively bounded analysis. This is not unexpected, given the fact that the truncation error becomes relevant for this
    type of extrapolation of the available LQCD data. It gives rise to concern for the theory prediction of non-local FFs,
    which are crucially reliant on this extrapolation. The biggest concern here is the FF $f_T^{B\to K}$, which shows deviations
    of the order of $10\%$.
\end{itemize}

\section{Orthonormal Polynomials}
\label{app:orthonormal_polynomials}

Our parametrization of the FFs involves orthonormal polynomials on the unit circle.
For the purpose of our analysis, orthonormality is defined with respect to the scalar product
\autoref{eq:th:scalar_product}, i.e., orthonormality with respect to integration over the arc
$\mathcal{A}_\alpha = \{e^{i\theta}, -\alpha < \theta < \alpha\}$.
We could not find a closed analytic formula for these polynomials in the literature. Instead, we evaluate them recursively
using the Szeg\H{o} recurrence relation~\cite{Simon2005}:
\begin{equation}
    \label{eq:simple_recurrence}
    \varphi_{n+1}(z) = z \, \varphi_n(z) - \bar{\rho}_n \, \varphi_n^{*,n}(z),
\end{equation}
where $\varphi_n$ is the $n$th orthogonal polynomial.
We adopt here the common notation used in the mathematical literature
$\varphi_n^{*,n}(z) \equiv z^n \bar{\varphi}_n(1/\bar{z})$~\cite{Simon2005}.
The coefficients $\rho_n$, known as the Verblunsky coefficients, uniquely define the set of polynomials and only depend on the integration measure.
In our approach the relevant integration measure is the constant Lebesgue measure on the arc $A_\alpha$.

Since this arc is symmetric under complex conjugation, the Verblunsky coefficients are real numbers~\cite{Simon2005}.
The orthonormal polynomials $p_n$ appearing in \autoref{eq:hFexp} are obtained using~\cite{Simon2005}
\begin{equation}
    p_n = \frac{\varphi_n}{||\varphi_n||}, \qquad 
    ||\varphi_n||^2 = 2 \pi \prod_{k=0}^{n-1}(1 - |\rho_k|^2),
\end{equation}
i.e. $||\varphi_0||^2 = 2\pi$, \textit{etc}\dots

\subsubsection*{Calculation of Verblunsky coefficients}
The Verblunsky coefficients can be calculated by applying the Gram-Schmidt orthonormalisation procedure to the basis of monomials $z^k$.
A faster evaluation can however be obtained recursively and does not require any analytic calculation.
To illustrate our approach to this evaluation, we first define $\Phi_n \equiv (\varphi_n(z), \varphi^{*,n}_n(z))^T$,
which fulfils the recurrence relation
\begin{equation}
    \label{eq:vector_recurrence}
    \Phi_{n+1} =
    \begin{pmatrix}
        z            & -\rho_n \\
        -z \, \rho_n & 1
    \end{pmatrix} \Phi_n.
\end{equation}
This equation can be used to evaluate the orthogonal polynomials at any point $z$, provided that the set of Verblunsky coefficient is known.
To compute the later, we introduce two sets of integrals,
\begin{equation}
    I_{n,k} \equiv \int_{-\alpha}^\alpha e^{ik\theta} \varphi_n(e^{i\theta}) d\theta \quad
    \mathrm{and} \quad
    J_{n,k} \equiv \int_{-\alpha}^\alpha e^{ik\theta} \varphi^{*,n}_n(e^{i\theta}) d\theta.
\end{equation}
For any non-zero integers $k,n$ one readily finds
\begin{equation}
    I_{0,0} = J_{0,0} = 2\alpha, \quad I_{0,k} = J_{0,k} = \frac{2\sin k\alpha}{k}, \quad I_{n,0} = 0.
\end{equation}
Integrating \autoref{eq:vector_recurrence} on the arc $\mathcal{A}_\alpha$ yields
\begin{equation}
    \begin{pmatrix}
        I_{n+1,k} \\ J_{n+1,k}
    \end{pmatrix} =
    \begin{pmatrix}
        1           & -\rho_{n} \\
        -\rho_{n} & 1
    \end{pmatrix}
    \begin{pmatrix}
        I_{n,k+1} \\ J_{n,k}
    \end{pmatrix}.
\end{equation}
The Verblunsky coefficient $\rho_m$ can then be obtained by recursively filling the triangular matrices $I_{n,k}, J_{n,k}$ for $k < n \leq m$
and using $\rho_m = I_{m,1} / J_{m,0}$.
This method is implemented in the open-source \EOS software \cite{EOSAuthors:2021xpv} used for numerical aspects of this paper.

\subsubsection*{Asymptotic behaviours}
The asymptotic behaviours of the orthonormal polynomials have been studied in details, see e.g. \cite{BH1998} and references therein.
Using Eq.~(1.7) of Ref.~\cite{Krasovsky2004} and applying the transformation $\theta' = \pi-\theta$, we find that at large $n$,
\begin{equation}
    \label{eq:verblunsky_large_n}
    \rho_n \sim (-1)^n \cos \frac{\alpha}{2}.
\end{equation}
This result can be used when diagonalising the matrix in \autoref{eq:vector_recurrence} to obtain asymptotic values for any $z \neq 0$.
In particular one finds that at large $n$,
\begin{equation}
    \label{eq:values}
    \frac{p_{n+2}(1)}{p_{n}(1)} \sim 1 \qquad
    \mathrm{and} \qquad
    \frac{p_{n+1}(-1)}{p_{n}(-1)} \sim -\cot\frac{\alpha}{4} < -1.
\end{equation}
The values of the polynomial thus form a finite alternating series at $z = 1$ and an alternating and exponentially divergent series at $z = -1$, $p_{n}(-1) = \mathcal{O}\left((\cot\frac{\alpha}{4})^n\right)$.

\subsubsection*{Convergence of the form factor expansion}
Following the approach of Ref.~\cite{Buck:1998kp}, we evaluate \autoref{eq:hFM} at the branch point $z(s_\Gamma) = -1$,
\begin{equation}
    \label{eq:ffatsgamma}
    \left. \left( \P_\F(z) \phi_\F(z) \FM{\Gamma,\lambda}(z) \right) \right|_{z = z(s_\Gamma)}
    =
    \sum_{n \geq 0} a^\F_n \, p_n^\F(-1)\,.
\end{equation}
As shown in \autoref{eq:values}, the series $p_n^\F(-1)$ diverges exponentially with $n$.
This, however, is not an issue for the parametrization. The power series element of the parametrization is
analytic for $|z| < 1$. Moreover, this element is also finite at $z = -1$. By Abel's theorem, the value
at $z = -1$ can be expressed as
\begin{equation}
    \sum_n a_n^{\F} p_n^{\F}(-1) < \infty\,.
\end{equation}
Together with \autoref{eq:values}, this implies that the coefficients $a_n$ must fall off at least exponentially to compensate the divergence of the series $\{p_n(-1)\}$.

\bibliographystyle{JHEP}
\bibliography{references}

\end{document}